\documentclass[journal]{IEEEtran}
\ifCLASSINFOpdf
\else
\fi
\usepackage{fixltx2e}
\usepackage{stfloats}

\usepackage{extarrows}
\usepackage{algorithm}
\usepackage{algpseudocode}
\usepackage{amsfonts}
\usepackage{amsmath}
\usepackage{graphicx}% Include figure files
\usepackage{dcolumn}% Align table columns on decimal point
\usepackage{bm}% bold math
\usepackage{algorithmicx}

\ifCLASSOPTIONcompsoc
\usepackage[caption=false,font=normalsize,labelfon
t=sf,textfont=sf]{subfig}
\else
\usepackage[caption=false,font=footnotesize]{subfi
g}
\fi

 \newtheorem{thm}{Theorem}
  \newtheorem{prop}{Proposition}
 
 \newtheorem{lem}[thm]{Lemma}
 \newtheorem{defn}{Definition}

\begin{document}

\hyphenpenalty=1000
\tolerance=500

\title{Quantum Radon Transform and Its Application}

\author{Guangsheng~Ma,
        Hongbo~Li,
        and~Jiman~Zhao, % stops a space
\thanks{G. Ma, corresponding author, is with Academy of Mathematics and Systems Science, Chinese Academy of Sciences, Beijing, 100875, China. e-mail: 627362183@qq.com.}
\thanks{H. Li is with Academy of Mathematics and Systems Science, Chinese Academy of Sciences; University of Chinese Academy of Sciences, Beijing, 100190, China. e-mail: hli@mmrc.iss.ac.cn.}% <-this % stops a space
\thanks{J. Zhao is with School of Mathematical Sciences, Beijing Normal University, Beijing, 100875, China. e-mail: jzhao@bnu.edu.cn.}}% <-this % stops a space
\maketitle

\begin{abstract}
This paper extends the Radon transform, a classical image processing tool for fast tomography and denoising, to the quantum computing platform. A new kind of periodic discrete Radon transform (PDRT), called quantum Radon transform (QRT), is proposed. The QRT has a quantum implementation that is exponentially faster than the classical Radon transform. Based on the QRT, we design an efficient quantum image denoising algorithm. The simulation results show that QRT preserves the good denoising capability as in the classical PDRT. Also, a quantum algorithm for interpolation-based discrete Radon transform (IDRT) is proposed, which can be used for fast line detection. Both the quantum extension of IDRT and the line detection algorithm can provide polynomial speedups over the classical counterparts.
\end{abstract}


% \fi
\IEEEpeerreviewmaketitle

\section{Introduction}
\IEEEPARstart{R}{adon} transform, proposed by Johann Radon in 1917 \cite{radon1917bestimmung}, is an important image processing tool with widespread applications in computed tomography, geophysics, and remote sensing, etc. \cite{deans2007radon}. It changes a function $f$ defined on the plane to a function $R_{\theta}f(\rho)$ defined on the space of lines in the plane, whose value on the line with interception $\rho$ and slope $\theta$ equals the integral of function $f$ along the line:
\begin{align}\label{x2}
R_{\theta}f(\rho):=\iint_{\mathbb{R}^2} f(x,y)\delta(\rho-x\sin\theta+y\cos\theta)dxdy,
\end{align}
where $\theta \in (0,\pi]$, $\rho\in\mathbb{R}$, and $\delta$ is the Dirac function. By definition, Radon transform possesses the capability of detecting singularities along straight lines, and performs better at denoising images with linear singularities than other image processing tools \cite{deans2007radon,do2000image}.

To implement Radon transform, discretization is necessary. However, the different discretization methods will result in different discrete Radon transforms (DRTs) that have different applications. In discretization, there are two methods to approximate the line integral in (\ref{x2}): the interpolation method, and the periodic discrete grid method.

The interpolation method (IDRT) evaluates the integral along a straight line by making interpolation among the adjacent points on the discrete image grid of the line. The earlier DRTs are based on this method \cite{radon1917bestimmung}. In 1987, Beylkin discovered an exact inversion formula, and proved that if the discrete version is based on Radon's original formula, then the reconstruction can only be approximate \cite{beylkin1987discrete}. The IDRT can be used in line detection, X-ray computed tomography \cite{toft1996radon}, etc. Performing the IDRT on an $N \times N$ image often requires at least $\Omega(N^3)$ arithmetic operations \cite{kelley1993fast}.

The periodic discrete grid method (PDRT) calculates the integrals along a set of warped lines, and does not have direct connection with the continuous Radon transform \cite{kelley1993fast,lun2017discrete,carranza2018fast}. Still the method possesses some very nice properties, such as the exact reconstruction property, Fourier slice property, etc. Matus and Flusser \cite{matus1993image} first investigated PDRT on $\mathbb{Z}^2_{p}$, where $p$ is prime. Then Hsung et al. \cite{hsung1996discrete} extended PDRT to $\mathbb{Z}^2_{p^n}$. The PDRT has been used in image denoising \cite{do2000image}, tomographic reconstruction \cite{svalbe2001reconstruction}, image watermarking and encryption \cite{kingston2006projective}, etc. To compute the PDRT of an $N\times N$ image, $\Omega(N^2 \text{log}N)$ arithmetic operations are required \cite{khanipov2018computational}.

With the emergence of quantum computing, for many important computational problems, it is found that quantum algorithms can provide dramatic speedup \cite{shao2019quantum}, for example, exponentially fast quantum algorithms such as quantum Fourier algorithm, Shor's factoring algorithm \cite{shor1999polynomial}; polynomially fast quantum algorithms such as Grover's search algorithm \cite{zalka1999grover}, and so on \cite{nielsen2000quantum}.

%\cite{ajtai1997public,ambainis2004quantum,laarhoven2013solving}.\cite{boneh1996algorithms,childs2003exponential}

%several quantum walk based algorithms \cite{boneh1996algorithms,childs2003exponential}, and some quantum algorithms on finite abelian hidden subgroup problem \cite{nielsen2000quantum}.

%For many other problems, quantum computation can only provide polynomial speedup \cite{ajtai1997public,ambainis2004quantum,laarhoven2013solving}, the most famous of which is Grover's search algorithm \cite{zalka1999grover}. For some other problems, however, quantum computation performs worse than classical computation, notably in preparing the overhead for each iteration step \cite{kerenidis2020quantum}.

To speed up image processing in the forthcoming quantum computing age, quantum image processing (QIMP), whose topics range from quantum image representations to image processing, has drawn a lot of attention in the last decade \cite{abura2017advances,beach2003quantum,yan2016survey}. Early work on QIMP concentrated on the quantum representation of images e.g., \cite{le2011flexible,li2014multi,venegas2003storing}. Roughly speaking, there are two typical representation methods: (1) amplitude representation method, such as the Real Ket representation \cite{latorre2005image} which utilizes the qubits' amplitude and a computational basis state to encode the grayscale and the location of a pixel respectively; (2) basis representation method, such as the novel enhanced quantum representation (NEQR) \cite{zhang2013neqr} which utilizes a register's computational basis state to encode both the grayscale and the location of pixels.

%Song et al. \cite{song2013dynamic} proposed a quantum image watermarking scheme using quantum wavelet transform.
%\cite{akhshani2012image}

%Quantum image preparation \cite{clader2013preconditioned,shao2019quantum,zhao2004experimental}, also known as input data preparation, has drawn a lot of attention in recent years.

%The first work on QIMP was done by in 1997 \cite{vlasov1997quantum}.

Another focus in QIMP is to develop image processing tools in the quantum computation framework \cite{fijany1998quantum,zhang2015qsobel}, for example, Yao \cite{yao2017quantum} proposed an efficient quantum image edge detection algorithm for Real Ket images. Image segmentation, watermarking, scrambling, and some other image processing problems have also been investigated in QIMP \cite{caraiman2014histogram,song2013dynamic,li2019block}.

However, there is still no extension of Radon transform to the field of quantum image processing. In this paper, we make such extension.

We first extend PDRT to quantum computation framework. The main difficulty in making the extension comes from the fact that there is no preference of using unitary transforms in designing classical algorithms; conversely, quantum algorithms mostly use unitary operators. To realize the classical `non-unitary' PDRT with unitary transformations, we design a quantum reversible multiplication, and then use Fourier slice property of PDRT with replacing the traditional multiplication by the quantum reversible multiplication. Finally, we obtain a transform that is different from any existing PDRTs. We call this new transform quantum Radon transform (QRT).

The QRT preserves many good properties as in the classical PDRT, such as Fourier slice property. By replicating the denoising experiments designed specially for testing PDRT \cite{do2000image}, it is shown that the QRT is of the good denoising capability as in the classical PDRT (cf. Fig. \ref{21}-\ref{fig:image10}). The most important advantage of QRT is that for an $N\times N$ image, the QRT can be implemented in time $O(\log^3 N)$, which runs exponentially faster than the classical PDRT, which has runtime $\Omega(N^2 \log N)$. As the application, a QRT-based quantum denoising method is proposed, which runs exponentially faster than the classical PDRT denoising method.

We then extend IDRT to the quantum case. The quantum extension of IDRT can provide polynomial speedup, with the input being a Real ket quantum image and the output being encoded with the NEQR representation. As the application, a quantum algorithm for line detection using IDRT is given, which enables the line detection process to be speeded up polynomially in the average case.

%Despite its adopted simple interpolation method, SIDRT is still useful enough that it can be used in line detection.

%This paper is arranged as follows. In Section $\ref{sec2}$, we first review some backgrounds on the Radon transform and then establish our main result. In Section $\ref{sec2.1}$, we present a reversible quantum multiplication, prepared for implementing the QRT in the following. In Section $\ref{sec2.2}$, we propose the definition and some properties of QRT. The relevant quantum algorithm and numerical experiment are also given in this section. In Section $\ref{sec3}$, we provide some results of extending interpolation-based DRT to the quantum case, and remain some potential quantum applications in the Section  $\ref{sec4}$.

This paper is arranged as follows. In Section \ref{secx1}, we introduce some background on Radon transform and classical/quantum image processing. In Section \ref{sec2.1}, we present a reversible quantum multiplication. In Section \ref{sec2.2}, we introduce QRT and explore some basic properties of it. In Section \ref{sec3}, we extend interpolation-based DRT to the quantum case. In Section \ref{sec4}, we present two applications of our proposed quantum transforms.

%which is follows immediately from the Fourier slice property.

%Although the wavelets and high-dimension wavelets have been discussed and extended in the quantum case.

\section{Preliminaries}\label{secx1}
\subsection{Periodic Discrete Radon Transform}\label{sec2}

Throughout this paper, we use $p$ to denote a prime number, use $n$ to denote a positive integer, and use $N$ to denote a power of $2$. $[n]$ is the subset of integers $\{0,1,...,n-1\}$. We use $I$ to denote the imaginary unit. The $L^2$-norm of vector $\vec{a}=(a_0,...,a_{n-1})$ is $||\vec{a}||_2=\sqrt{\sum_{i\in[n]} |a_i|^2}$.

\begin{figure}[!t]
\hspace{5mm}\includegraphics[width=2.5in]{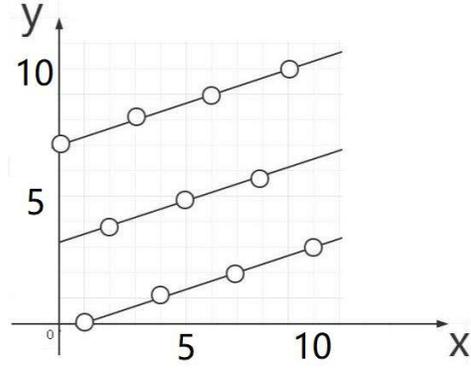}
\caption{Example of a discrete line $L^{11}_{1,3}=\{(x,y)|x-3y=1 \text{ mod } 11\}$ on lattice $\mathbb{Z}^{2}_{11}$. The hollow circles indicate the lattice points over the discrete line $L^{11}_{1,3}$.}
\label{9}       % Give a unique label
\end{figure}

We begin with a specific kind of discrete `line', an example of which is given in Fig. \ref{9} :

\textbf{Discrete line $L^{n}_{l,k}$}. The discrete line on lattice $\mathbb{Z}^{2}_{n}$ with interception $l$ and slope $k$ is
\begin{equation}\label{1}
L^n_{l,k}=\left\{
\begin{aligned}
&\{ (x,y)\ \big{|} x+ky=l\mod l, \ x, y\in[n]\}, \text{if}\ k\in[n]; \\
&\{ (x,l) \ \big{|} x\in[n]\}, \quad \text{if}\quad k=n.
\end{aligned}
\right.
\end{equation}

The periodic discrete Radon transform (PDRT) -- sometimes called finite Radon transform -- is defined as summations of function values at points over these discrete lines.

\begin{defn}[Periodic discrete Radon transform \cite{matus1993image}]\label{1}
The PDRT of a function $f$ defined on lattice $\mathbb{Z}^{2}_{n}$ is
\begin{align}\label{223}
r_{k}(l)=\frac{1}{\sqrt{n}}\displaystyle\sum_{(i,j)\in L^{n}_{l,k}} f(i,j), \quad l\in[n], \ k\in[n+1].
\end{align}
\end{defn}

The following proposition gives an important property of PDRT:
\begin{prop}[Fourier slice property of PDRT \cite{do2000image}]\label{xx3}
Let $\mathcal{F}_1$ and $\mathcal{F}_2$ be the $1$-D and $2$-D discrete Fourier transform, respectively. For a function $f$ defined on $\mathbb{Z}^{2}_{n}$, let $r_k(l)$ be the PDRT of $f$, so that it is defined on $\mathbb{Z}_{n}$. Then for any $\omega, k \in[n]$,
\begin{align}\label{tra}
\mathcal{F}_1\{r_{k}\}(\omega)=\mathcal{F}_2\{f \}(\omega, k\omega \emph{ mod } n);
\end{align}
for $k=n$,
\begin{align}
\mathcal{F}_1\{r_{n}\}(\omega)=\mathcal{F}_2\{f \}(0,\omega).
\end{align}
\end{prop}
\begin{IEEEproof}
For any $k\in[n]$, the discrete Fourier transform of function $r_k(l)$ in variable $l$ is
\begin{eqnarray}
\mathcal{F}_1\{r_{k}\}(\omega)&=&\frac{1}{n}\displaystyle\sum_{l,y\in[n]}f(l-ky,y)e^{-2 \pi I \frac{l\omega}{n}}\nonumber\\
    &=&\frac{1}{n}\displaystyle\sum_{l,y\in[n]}f(l-ky,y)e^{-2 \pi I \frac{(l-ky)\omega+yk\omega}{n}}\nonumber\\
    &=&\mathcal{F}_2\{f\}(\omega,k\omega \text{ mod } n).
\end{eqnarray}
For $k=n$,
\begin{align}
\mathcal{F}_1\{r_{n}\}(\omega)&=\frac{1}{n}\displaystyle\sum_{x,l\in[n]}f(x,l)e^{-2 \pi I \frac{l\omega}{n}}=\mathcal{F}_2\{f\}(0,\omega).
\end{align}
\end{IEEEproof}

This Fourier slice property provides a fast implementation of PDRT. For an $n\times n$ image $f$, each value of its PDRT can be computed in time O$(n)$, so the whole PDRT can be obtained in time O$(n^3)$ if directly computed by definition. On the other hand, the Fourier slice property of PDRT allows one to compute the PDRT in time O$(n^2 \text{log} n)$: since the $1$-D and $2$-D (inverse) Fourier transforms can be implemented in time O$(n \text{log} n)$ and O$(n^2 \text{log} n)$ \cite{nussbaumer1981fast}, and the PDRT of $f$ can be obtained by performing $1$-D inverse Fourier transform on $\mathcal{F}_2\{f\}$ according to (\ref{tra}), the computational complexity of the PDRT is thus reduced to O$(n^2 \text{log} n)$.

Besides designing fast algorithms, another important practical issue is to recover the original image from its Radon transform. In \cite{matus1993image}, Mat$\acute{u}\breve{s}$ and Flusser proposed the following reconstruction formula for PDRT on $\mathbb{Z}^2_p$:

\begin{prop}[Reconstruction formula for PDRT on $\mathbb{Z}^2_p$]
Let $r_{k}(l)$ be the PDRT of a function $f$ defined on $\mathbb{Z}^2_p$. Then for any $i, j \in [p]$,
\begin{align}\label{x1}
f(i,j) =\frac{1}{\sqrt{p}}\sum_{\substack{\{ (l,k)|(i,j)\in L^{p}_{l,k},\\ l\in[p],\ k\in[p+1] \}}}  r_{k}(l)-\frac{1}{p}\displaystyle\sum_{x,y\in[p]}f(x,y).
\end{align}
\end{prop}

\begin{IEEEproof}
The following geometric properties of discrete lines are easy to verify:
\begin{itemize}
\item[1.] Every discrete line $L^{p}_{l,k}$ contains $p$ lattice points, and two parallel discrete lines have no point of intersection;
\item[2.] For any fixed slope $k$, the $p$ parallel lines $L^{p}_{l,k}$ (where $l\in[p]$) provide a complete cover of the lattice $\mathbb{Z}^{2}_{p}$;
\item[3.] Two discrete lines $L^{p}_{l,k}$ of different slopes will interact in exactly one point.
\end{itemize}
For all the $p+1$ lines through a fixed point $(i,j)$, by term 1, every two of them have only one point in common, which is just $(i,j)$. Since there are $p(p+1)$ points on the lines, and there are $p+1$ copies of point $(i, j)$ on such lines, there are all together $p^2$ different points on these lines, which are exactly the total number of points in lattice $\mathbb{Z}_{p}^2$:
\begin{align}\label{vcb1}
\displaystyle\sum_{x,y\in[p]}f(x,y)+&pf(i,j)=\sqrt{p}\sum_{\substack{  \{ (l,k)|(i,j)\in L^{p}_{l,k},\\ l\in[p],\ k\in[p+1] \}  }} r_{k}(l).
\end{align}
\end{IEEEproof}
%By (\ref{vcb1}), one can exactly recover an image $f$ from its PDRT projections if the mean of $f$ is $0$, or has been subtracted from $f$ before taking PDRT.

In general, the size of an image is not the square of a prime number. There is a series of work to extend PDRT to images of more general sizes \cite{hsung1996discrete,kingston2007generalised,matus1993image}. For one example, Kingston \cite{kingston2006orthogonal} extends PDRT to images of size $p^n \times p^n$. The reconstruction formula for such images are much more complicated, e.g., (11) in \cite{kingston2006orthogonal}.

%It is the reconstruction property that enables PDRT to be used in image denoising (see \cite{do2000image} for more details).

\subsection{Classical Image Denoising}\label{41}
Image denoising is to remove noise from a noisy image, so as to restore the true image \cite{buades2005review,donoho1994ideal}. Suppose we are given a real-valued noisy signal
\begin{align}
h_i=f_i+e_i, \quad i\in [n],
\end{align}
where $\vec{f}=(f_0,f_1,...,f_{n-1})$ is original signal, and $e_i$ is the noise sampled independently from the normal distribution $V(0,\sigma^2)$, where $0$ is the mean and $\sigma^2$ is the variance. After performing some denoising method on $\vec{h}=(h_0,h_1,...,h_{n-1})$ to prepare the proceed signal $\vec{h'}$, if the following noise level decreases (i.e., $\text{Risk}(\vec{h},\vec{f})>\text{Risk}(\vec{h'},\vec{f})$):
\begin{align}\label{bm}
\text{Risk}(\vec{h},\vec{f})=\frac{1}{n} E(||\vec{h}-\vec{f}\ ||^2_2),
\end{align}
where $E(\cdot)$ is the expectation, then we say such denoising method is effective.

\begin{figure}
\includegraphics[scale=0.14]{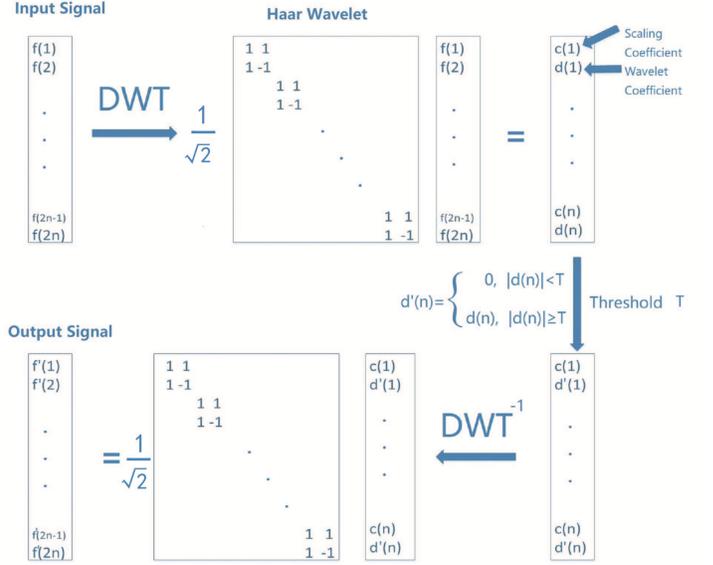}
\caption{Denoising using DWT. There are three steps in total: (1) Apply the DWT to the input noisy signal $f$, i.e., multiply $f$ by the matrix form of some wavelet (the one used here is Haar wavelet). The resulting coefficients can be divided into two parts: the scaling coefficients $c$, composed of the odd rows, and the wavelet coefficients $d$, composed of even rows. (2) Apply Hard-thresholding to wavelet coefficients $d$ with threshold $T$ to obtain new wavelet coefficients $d'$ (3) Apply inverse DWT to thresholded coefficients, where in this case the inverse of the Haar wavelet matrix is itself. }
\label{xfig:2}       % Give a unique label
\end{figure}

Fig. \ref{xfig:2} shows a general procedure for denoising signal using discrete wavelet transform (DWT). Below, we explain why DWT denoising method is effective. Let $\vec{f}=(f_i)$ be a pure signal where $f_i\equiv1$, $i\in [n]$. Let $\vec{h}=(h_i)$ be the noisy signal where $h_i=1+e_i$, $e_i \sim V(0,\sigma^2)$. After denoising $\vec{h}$ using Haar wavelet and threshold $T=\infty$ (i.e, change all wavelet coefficients to $0$), by Fig. \ref{xfig:2}, each element of denoised signal is of the form $h'_i=1+e'_i$, where
\begin{align*}
e'_i \sim \frac{1}{2}V(0,\sigma^2)+\frac{1}{2}V(0,\sigma^2)=V(0,\sigma^2/2).
\end{align*}

Now that $\text{Risk}(\vec{h'},\vec{f})=\frac{\sigma^2}{2}<\frac{\sigma^2}{n}=\text{Risk}(\vec{h},\vec{f})$, the noise level decreases, and thus the Haar denoising method works.

However, for general denoising algorithms, it is often hard to make such a statistical analysis as the above. Experimentally, the following so-called signal-noise-ratio (SNR) is often taken to measure denoising performance:
\begin{align}\label{snr}
\text{SNR}(\vec{h},\vec{f}) =10 \log_{10}(\frac{||\vec{h}||_2^2}{||\vec{h}-\vec{f}||^2_2 } ),
\end{align}
where $f$ and $h$ are some specific test signals. In comparison with the probability value in (\ref{bm}), SNR is more easily accessible by numerical experiments. A denoising method is said to be effective if $\text{SNR}(\vec{h'},\vec{f})>\text{SNR}(\vec{h},\vec{f})$.

Now, we consider image denoising, i.e., $2$-dimensional signal denoising. The simplest $2$-D DWT denoising method is to apply $1$-D denoising method to each row in the signal matrix, then use $1$-D denoising method to each column. However, image possesses various geometric features. By the work of Do and Vetterli \cite{do2000image}, the following PDRT denoising method are more effective than the general $2$-D DWT in denoising
images with obvious singularities along straight lines, as shown in Fig. \ref{21}.

\begin{itemize}
\item[1)] Apply the PDRT to the noisy image to obtain its PDRT $r_k(l)$,

\item[2)] For each slope $k$, perform once 1-D DWT denoising method on the PDRT $r_k(l)$ along the direction of interception $l$.

\item[3)] Perform inverse PDRT by (\ref{vcb1}).
\end{itemize}

In \cite{do2000image}, the authors explain why PDRT is better: by PDRT, the typical linear singularities of pure image are represented by a few wavelet coefficients (in step 2) while randomly located noisy singularities are unlikely to produce significant coefficients. This is unlike using the $2$-D DWT where both noisy pixels and image singularities can produce significant wavelet coefficients. Therefore one can remove the noise with less damage to the original image by properly thresholding wavelet coefficients.

%So, PDRT is more effective in denoising the image (function) that is smooth piecewise with singularities concentrated mainly on a straight line, such as Fig. \ref{21}.

A routine computation shows that the time complexity of donising an $n\times n$ image using $2$-D DWT is O$(n^2)$. The time complexity of PDRT denoising method is O$(n^3)$, because performing inverse PDRT has a time complexity O$(n^3)$ by (\ref{vcb1}).

\subsection{Quantum Image Representation and Preparation}\label{sec34}
\begin{defn}[Real ket representation \cite{latorre2005image} \label{5}] Let $f$ be an $N \times N$ image, where $N=2^n$, and $f(i,j)\geq 0$ is the image intensity at point $(i,j)$, then its Real Ket representation is
\begin{align}\label{vc14}
\frac{1}{\sqrt{\displaystyle\sum_{i,j\in[N]} |f(i,j)|^2}} \sum_{i,j\in[N]} f(i,j)|i\rangle|j\rangle,
\end{align}
\end{defn}

\begin{defn}[Novel enhanced quantum representation \cite{zhang2013neqr} \label{77}]The NEQR representation of image $f$ is
\begin{align}\label{vc4}
\frac{1}{N}\sum_{i,j\in[N]}|f(i,j)\rangle|i\rangle|j\rangle.
\end{align}
\end{defn}

In this paper, we use the Real Ket representation of images in the quantum computation framework. The famous quantum Fourier transform \cite{nielsen2000quantum} and HHL algorithm \cite{harrow2009quantum} are both based on this amplitude encoding method. Unless otherwise specified, the term `quantum image' refers to an image encoded with the Real Ket method.

%The Real Ket is a kind of amplitude encoding method, based on which a number of quantum algorithms have been constructed, notably quantum Fourier transform algorithm and the HHL algorithm \cite{harrow2009quantum}.

To show the connection between the above two representation methods, we need an efficient operation called \emph{conditional rotation} \cite{harrow2009quantum,kerenidis2017quantum2}:
\begin{prop}[Conditional rotation]\label{218}
Let $a$ be the $k$-bit finite precision representation of a positive number that is smaller than 1. Then the following mapping can be performed in time O\emph{(}$k^{2.5}$\emph{)}:
\begin{align}
|a\rangle|0\rangle\rightarrow|a\rangle(a|0\rangle+\sqrt{1-a^2}|1\rangle).
\end{align}
\end{prop}
\begin{IEEEproof}
Let $a:=\displaystyle\sum_{i \in[k]} 2^{-i-1}a_{i}$, where $ a_i \in \{ 0,1 \}$. Then $|a_i\rangle$ is the $(i+1)$-st qubit of $|a\rangle=|a_{k-1}\rangle\cdots|a_1\rangle|a_0\rangle$ counted from right. For $s>0$, define 1-qubit quantum gate
\begin{equation}       % 开始数学环境
R_{s}=\left[                 %左括号
  \begin{array}{cc}   % 该矩阵一共3列，每一列都居中放置
\cos s,& -\sin s\\
\sin s, &\cos s
  \end{array}
\right].     %右括号
\end{equation}
Then it holds that
\begin{equation}
\prod_{i\in[k]} R^{\alpha_i}_{2^{-(i+1)}}=\prod_{i\in[k]} R_{2^{-(i+1)}\alpha_i}=R_{\alpha}.
\end{equation}
By performing the following $k$ successive 1-bit conditional rotations: for each $i\in[k]$ the corresponding rotation is $R^{\alpha_i}_{2^{-(i+1)}}$ where $\alpha_j\in\{0,1\}$ is the control bit, on the last qubit of $|a\rangle|0\rangle$, one gets

\begin{equation}\label{2}
|a\rangle|0\rangle\rightarrow |a\rangle (\cos(a)|0\rangle+\sin(a)|1\rangle).
\end{equation}

By Lemma $48$ in \cite{gilyen2019quantum},  given $|a\rangle$ where $a\in(0,1)$, using Taylor series approximation allows to prepare the state $|\arccos(a)\rangle$ in time O$(k^{2.5})$, where $\arccos(a)\in[0,\frac{\pi}{2})$. So with the help of ancilla qubits and (\ref{2}), the following sequence of mappings can be implemented in time O$(k^{2.5})$:
\begin{align*}
|a\rangle|0\rangle|0\rangle &\longrightarrow |a\rangle|\arccos(a)\rangle|0\rangle\\
&\longrightarrow|a\rangle|\arccos(a)\rangle(a|0\rangle+\sqrt{1-a^2}|1\rangle)\\
&\longrightarrow|a\rangle|0\rangle(a|0\rangle+\sqrt{1-a^2}|1\rangle).
\end{align*}
\end{IEEEproof}

By the following proposition, a Real Ket image can be prepared from its NEQR version, and this preparation procedure is efficient if the condition number $\kappa:=\frac{\min_{i,j\in[N]}|f(i,j)|}{\max_{i,j\in[N]}|f(i,j)|}$
is large enough.

\begin{prop}\label{6}
Let $\vec{a}$ be a real vector realized by a unitary operator $U:|i\rangle|0\rangle\rightarrow|i\rangle|a_i\rangle$, $i\in[N]$ where $a_i$ is the $m$-bit finite precision representation of the vector entries, and $U$ can be performed in time O$(T_U)$. Given $a_{max}=\max_{j}|a_j|$, the state $|\vec{a}\rangle$ can be prepared in time O$(\frac{T_U+\emph{poly}(m, \log N))}{\kappa^2})$ by the following mapping:
\begin{align}
|0\rangle\rightarrow \cos\theta|\vec{a}\rangle|0\rangle+\sin\theta |\vec{a}^{\perp}\rangle|1\rangle,
\end{align}
where

1) $|\vec{a}\rangle:=\displaystyle\sum_{i\in[N]} \frac{a_i}{||\vec{a}||_2}|i\rangle$;

2) $\vec{a}^\perp$ denotes the vector of entries $a^{\perp}_{i}:=\sqrt{1-|\frac{a_i}{a_{max}}|^2}$ for $i\in [N]$;

3) the coefficient $\cos\theta= \sqrt{\frac{||\vec{a}||^2_2  }{N a^2_{\max}}}\geq \frac{\min\limits_{i\in[N]} |a_i|}{\max\limits_{i\in[N]} |a_i|}:=\kappa $, and $\kappa$ is called the condition number of vector $|a\rangle$.
\end{prop}

\begin{IEEEproof}
With the number $a_{max}$ and unitary operator $U$ at hand, we can implement the following transform \cite{harrow2009quantum}:
\begin{align}\label{vx1}
|i\rangle|a_i\rangle\rightarrow|i\rangle|\frac{a_i}{a_{max}}\rangle.
\end{align}
We first perform the this transform on the input state $\displaystyle\sum_{i \in [N]} \frac{1}{\sqrt{N}}|i\rangle|a_i\rangle$ and then use the rotation conditioned on $|\frac{a_i}{a_{max}}\rangle$ (cf. Proposition \ref{218}):
{\small
\begin{align}
|i\rangle|\frac{a_i}{a_{max}}\rangle|0\rangle\rightarrow|i\rangle|\frac{a_i}{a_{max}}\rangle(\frac{a_i}{a_{max}}|0\rangle+\sqrt{1-(\frac{a_i}{a_{max}})^2}|1\rangle),
\end{align}}and finally undo $U$ to clear the second register. After the above operations, we measure the last qubit.

The running time of the above procedure is O$(T_{U}+\text{poly}(m, \log N))$. The expected result $|0\rangle$ by measurement has probability $\frac{\sum_{i\in[N]} a^2_i}{N a^2_{max}}\geq\frac{a^2_{min}}{a^2_{max}}=\kappa^2 $, which indicates that we have successfully prepared the state $|\vec{a}\rangle:=\displaystyle\sum_{i\in[N]} \frac{a_i}{||\vec{a}||_2}|i\rangle$.
\end{IEEEproof}

\emph{Remark:} It is also possible to convert a Real Ket image to its NEQR version by applying phase estimations \cite{nielsen2000quantum}. The close connections between different image representations give researchers more confidence in developing quantum image processing tools based on a particular representation, as they are likely to become universal once quantum techniques are sufficiently developed.

Quantum image preparation, also known as quantum (initial) state preparation, has been extensively studied over years, e.g., \cite{biamonte2017quantum,clader2013preconditioned,kerenidis2017quantum2}.
Current techniques allow to efficiently prepare the Real Ket (or NEQR) state of an $N\times N$ image in time $O(\text{polylog}N)$ if the prepared image has some special structures\footnote{such as the corresponding quantum data structure \cite{kerenidis2017quantum2} is given or the quantum state to be prepared has a well condition number \cite{shao2018linear}.}. The time required to prepare an arbitrary quantum image is at most no more than the classical preparation time, up to a logarithmic factor. Specifically, a quantum image of form (\ref{vc14}) or (\ref{vc4}) can be prepared by applying at most $N^2$ conditional operations, each of which has a time complexity polylog($N$).

The following proposition will be used in Section \ref{sec3}. It states that the inner products of quantum states can be estimated in parallel:
\begin{prop}[Parallel swap test \cite{shao2018quantum2} \label{32}]. Given $2N$ quantum states $|\vec{u}_0\rangle$, $|\vec{v}_0\rangle...|\vec{u}_{N-1}\rangle$, $|\vec{v}_{N-1}\rangle$, and two state preparation unitaries: $|k\rangle|0\rangle\rightarrow|k\rangle|\vec{u}_k\rangle$ and $|k\rangle|0\rangle\rightarrow|k\rangle|\vec{v}_k\rangle$ (where $k\in[N]$) that can be implemented in time O$(T_{in})$. There is a quantum algorithm with runtime O$(\frac{T_{in}}{\epsilon})$ to achieve
$|i\rangle|0\rangle \rightarrow|i\rangle|s_i\rangle$ for all $i\in[N]$, where $|s_i-\langle \vec{u}_i|\vec{v}_i \rangle| \leq  \epsilon$.
\end{prop}

A detailed proof can be found in Theorem 1 in \cite{shao2018quantum2}.

\section{Quantum reversible multiplication}\label{sec2.1}
This section presents a reversible modular multiplication in the quantum computation framework. Fix $N=2^n$. For fixed $0\leq i<2^{n-1}$ and $C=2i$, the multiplication in $\mathbb{Z}_{N}$ by $C$ is irreversible, because the mapping: $a\rightarrow aC \mod N$ maps both $a=0$ and $a=2^{n-1}$ to $0$. On the other hand, for $D = 2i+1$, the multiplication in $\mathbb{Z}_{N}$ by $D$ is reversible, because if $aD=bD \text{ mod } N$ for some $a,b \in [N]$ and $a\neq b$, then $(a-b)D\mid 2^{n}$. Since $D$ is odd, it must be that $a - b$ is a multiple of $2^n$; in particular, $|a-b|\geq 2^{n}$, which is impossible for $a,b \in [N]$.

Below, we design a unitary realization of the
multiplication in $\mathbb{Z}_{N}$ by any odd number $D \in [N]$. For
$1\leq k \leq n$,
%Now, we will prove that the $2n$-qubit quantum gate $M^{n}$ performs modular multiplication for all odd $i \in[2^{n}]$ and positive integers $n$
\begin{itemize}
\item let $M_k$ be a to-be-realized unitary operator performing
\begin{align}\label{2134}
|a_k\rangle|b_k\rangle\rightarrow |a_k\rangle|a_k b_k\ \text{mod}\ 2^k\rangle, \forall a_k, b_k \in[2^k]\ \text{and odd}\ a_k.
\end{align}

\item Let $A_k$ be the following controlled addition in $\mathbb{Z}_{2^k}$:
\end{itemize}

\begin{equation*}
|a_k\rangle|b_k\rangle|c\rangle\rightarrow \left\{
\begin{aligned}
&|a_k\rangle|a_k+b_k \text{ mod } 2^{k}\rangle|c\rangle, &\text{if } c=1;\\
&|a_k\rangle|b_k\rangle|c\rangle, &\text{if } c=0;
\end{aligned}
\right.
\end{equation*} $\hfill \forall a_k, b_k \in [2^{k}], c \in [2].$

The time complexity for performing each $k$-qubit addition is O$(k^2)$ \cite{ruiz2017quantum}, so is the time complexity of performing unitary operator $A_k$ (see Section 4.2 in \cite{nielsen2000quantum}).

For any $1 \leq k \leq n$, for any $a_k \in [2^k]$, let the binary representation of integer $a_k$ be
\begin{align*}
B_{i_{k-1}...i_0}:=\displaystyle\sum_{l\in[k]}2^{l}i_{l},\quad \text{where}\ i_l \in [2].
\end{align*}

When $k = 1$, multiplication operator $M_1$ is the identity:
\begin{align*}
|1\rangle|b_1\rangle\rightarrow|1\rangle|b_1\rangle, \quad \forall\ b_1 \in[2].
\end{align*}

In the following, we realize multiplication operator $M_{k+1}$ by $A_l$ for $1\leq l \leq k$ recursively.

For any $a_{k+1},b_{k+1} \in [2^{k+1}]$ where $a_{k+1}$ is odd, let their binary representations be $B_{i_{k}...i_{1}1}$ and $B_{j_{k}...j_1j_0}$ respectively. Then their modular multiplication is
\begin{align}\label{8}
&(2^{k}i_k+B_{i_{k-1}...i_{1}1})(2B_{j_k...j_1}+j_0) \text{ mod } 2^{k+1} \\
=&(2B_{i_{k-1}...i_11}B_{j_{k}...j_1}+B_{i_{k}...i_{1}1}\frac{1+(-1)^{j_0-1}}{2}) \text{ mod } 2^{k+1} \nonumber\\
=&\big{\{} 2(B_{i_{k-1}...i_11}B_{j_{k}...j_1} \text{ mod } 2^{k})+2\delta^{1}_{j_0}B_{i_{k}...i_{1}}+j_0 \big{\}}\text{mod } 2^{k+1}\nonumber
\end{align}
where $\delta^{1}_{j_0}$ is the Kronecker symbol.

\begin{figure}
\hspace{5mm}\includegraphics[scale=0.3]{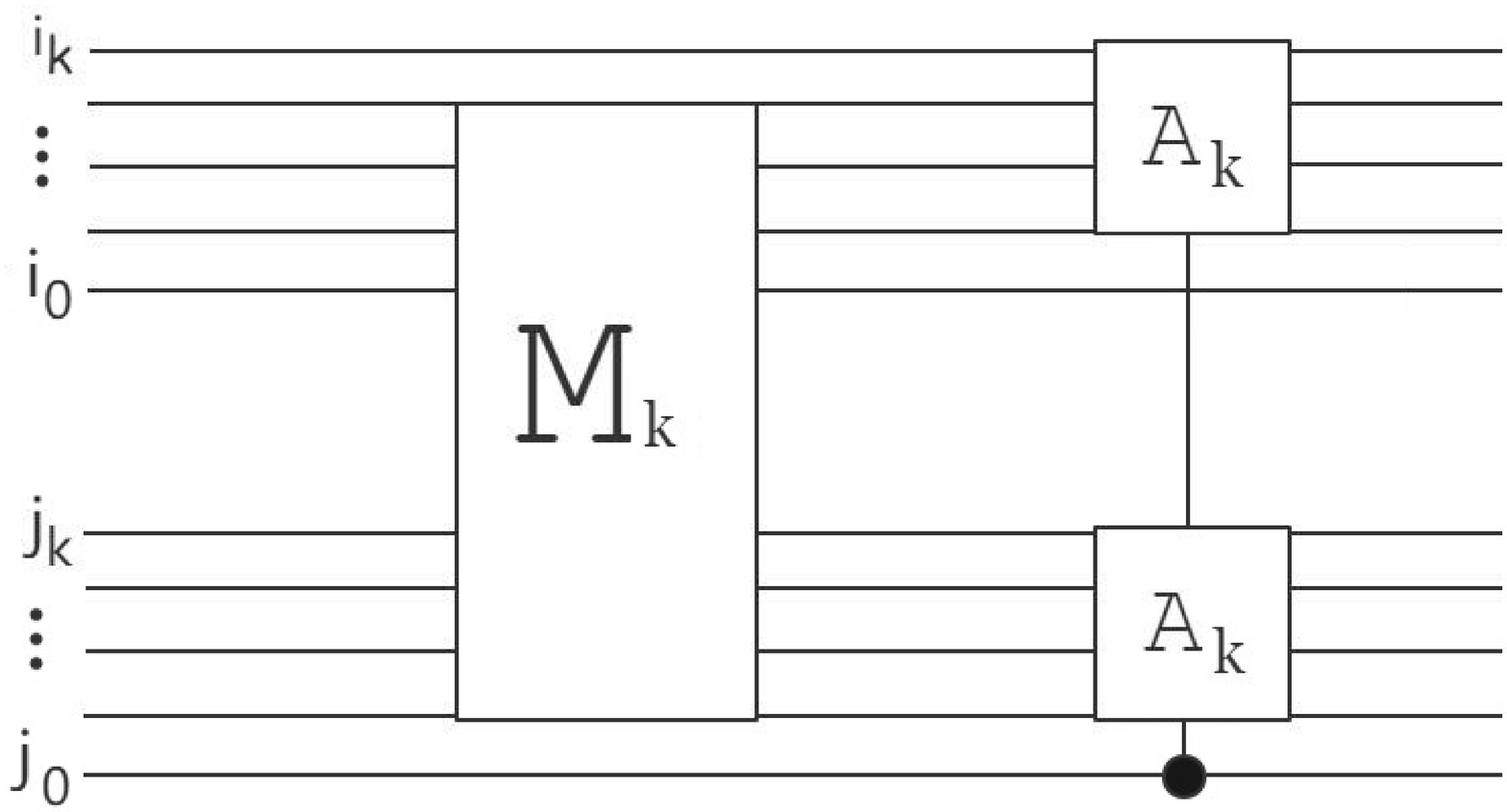}
\caption{The quantum circuit realizing $M_{k+1}$ by $M_{k}$ and $A_k$.}
\label{xfig:1}       % Give a unique label
\end{figure}

So the modular multiplication $M_{k+1}$ of two integers $B_{i_{k}...i_{1}1},B_{j_{k}...j_1j_0} \in [2^{k+1}]$, each containing $k+1$ binary digits, can be decomposed into two operators: the modular multiplication $M_k$ of integers $B_{i_{k-1}...i_{1}1},B_{j_{k}...j_1} \in [2^{k}]$, which occurs in the first $k$ binary digits, followed by the controlled modular addition $A_k$ of the above modular multiplication result and $B_{i_{k}...i_{1}}$, which also occurs in the first $k$ binary digits, while the control digit $j_0 \in [2]$ remains in the last binary digit. The quantum circuit realizing this decomposition is shown in Fig. \ref{xfig:1}.

In (\ref{8}) the modular multiplication $M_k$ can be further decomposed into $M_{k-1}$ and $A_{k-1}$, and by doing so recursively, $M_{k+1}$ is finally decomposed into a series of controlled modular additions: $A_1,A_2,\cdots,A_k$. Since performing $A_j$ has time complexity C$j^2$ for fixed constant $C>0$ and varying $1\leq j \leq k$, by $1^2+2^2+...+k^2=O(k^3)$, we get that the time complexity of realizing $M_{k+1}$ (hence $M_k$) by (14), is O$(k^3)$.

\section{Quantum Radon Transform}\label{sec2.2}

%We will present the definition of the QRT on $\mathbb{Z}^2_{n}$, where $n$ is a positive integer, then give the effective quantum implementation of the forward and inverse QRT on $\mathbb{Z}^2_{N}$, where $N$ is a power of $2$.
The classical PDRT in (\ref{223}) can be viewed as a mapping $\mathbb{R}^{n^2}\rightarrow\mathbb{R}^{n(n+1)}$ as follows:
\begin{align}
\Big{(}f(0,0),f(1,0),...,&f(n-1,n-1)\Big{)}\xrightarrow{\textnormal{PDRT}} \nonumber\\
 &\Big{(}r_0(0),r_0(1),...,r_n(n-1)\Big{)},
\end{align}
whose transformation matrix is not unitary\footnote{The lines $L^{n}_{0,0}$, $L^{n}_{0,1}$ have intersection, so the rows of transformation matrix corresponding to $r_0(0)$, $r_1(0)$ are not orthogonal.}.
It is hard to directly design a quantum algorithm for PDRT by (\ref{223}). So, we consider utilizing the related Fourier slice property in (\ref{tra}).

%Inspired by the case of fast/quantum Fourier transform \cite{nielsen2000quantum}, where the property that originally plays a key role in (classical) fast algorithm helps a lot in the quantum generalization,

By replacing the traditional multiplication in (\ref{tra}) with the quantum reversible multiplication, we successfully design an efficient quantum algorithm that can realize a novel transform similar to PDRT. We name it the quantum Randon transform:

%As far as we know, our quantum Radon transform is not just a quantum analogy of any existing DRTs. Here we first define this new kind of DRT in a classical way, instead of using the language of quantum computation.
\begin{defn}[Quantum Radon transform]\label{1021}
Let $f$ be a function defined on $\mathbb{Z}^{2}_{n}$, let $\lfloor \cdot \rfloor$ be the floor function, and let $\tilde{f}$ defined on $\mathbb{Z}^{2}_{2n}$ be related to $f$ as
{\small
\begin{align}\label{fstar}
\tilde{f}(x',y'):=\frac{1}{2} (-1)^{\lfloor \frac{x'}{n}\rfloor+\lfloor \frac{y'}{n}\rfloor} f  (  x' \text{\ mod } n,  y' \text{ mod } n  ), x',y'\in [2n].
\end{align}}Recall that $L^{2n}_{l,k}$ represent discrete lines on the lattice $\mathbb{Z}^{2}_{2n}$:
\begin{align}
L^{2n}_{l,k}:=\{ (x',y')|x'+ky'=l\ ( \text{mod} \ 2n),\ x',\ y'\in[2n] \}.\nonumber
\end{align}
Then, the \textbf{quantum Radon transform} of $f$ is a function defined on $\mathbb{Z}^{2}_{2n}$ as
\begin{align}\label{021}
QR_f(l,k)=\frac{1}{\sqrt{2n}}\displaystyle{\sum_{(x',y') \in L^{2n}_{l,k}  }}\tilde{f}(x',y'), \quad l,k \in[2n]
\end{align}
\end{defn}

The definition of QRT is derived from the following Algorithm \ref{xradon} rather than a deliberate construction, and it may seem a little complicated. Now, let us give a closer observation of $\tilde{f}$ in (\ref{fstar}). For any $x,y \in[n]$,
\begin{align}\label{cd}
f(x,y)&=\tilde{f}(x,y)=-\tilde{f}(x+n,y)=-\tilde{f}(x,y+n)\nonumber\\
&=\tilde{f}(x+n,y+n).
\end{align}
Graphically speaking, the QRT of an image $f$ can be viewed as the classical PDRT that performs on the symmetrized double-sized original image. Fig. \ref{x3} shows the particular summation method adopted by the QRT. Intuitively, the alternating sums seems to tend to suppress the changes in the Radon domain along the direction of interception, resulting in a lower sensitivity of QRT to linear singularities. Even worse, only a half of slopes (lines) could be detected by QRT, since when slope $k$ is even, by (\ref{cd}) it holds that
\begin{align}
QR_f(l,k)=0.
\end{align}

%offset against each other, and thus will

However, by experiments (see Fig. \ref{21}-\ref{fig:image10}), it has been found that QRT preserves the good denoising capability as in the classical PDRT; it will be further discussed in Section \ref{sec4.1} later. We hold the view that the QRT has a practical value comparable to that of the classical PDRT. On the other hand, as an advantage, QRT is invertible, which is derived from the following invertible QRT algorithm. The most important advantage lies in that the QRT can be performed exponentially faster than the classical PDRT, by the following quantum Radon transform algorithm:

\begin{figure}
\centering
\subfloat[Sum the values of $f$ at all white points.]{\includegraphics[width=1.4in]{zxian3.eps}\label{57}}
\hfil
\subfloat[Sum the values of $f$ at white points then subtract the values at black points.]{\includegraphics[width=1.4in]{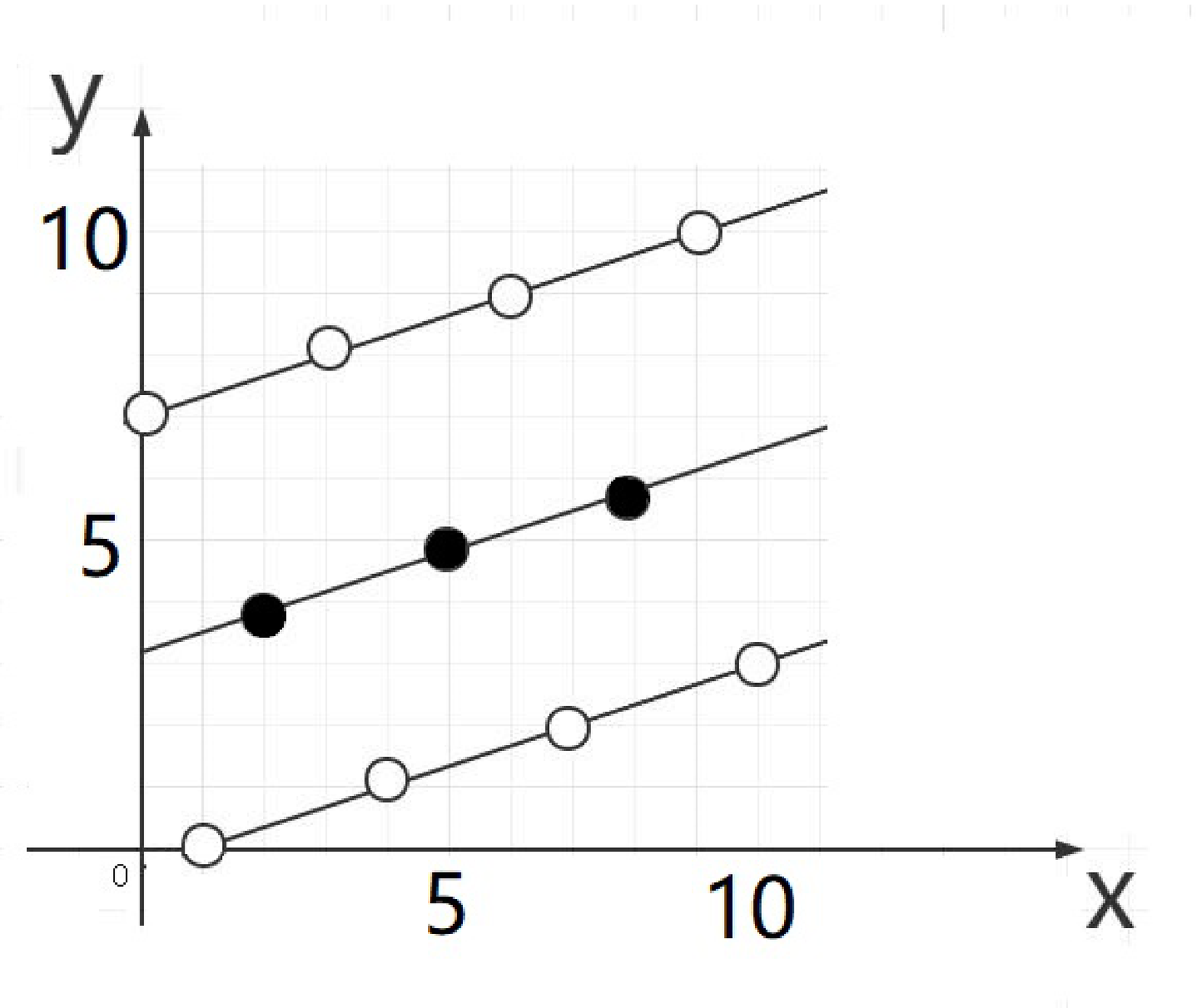}\label{56}}
\caption{A toy example to show the different summation methods adopted to compute the PDRT (left) of $f$ and QRT (right) of $f$, where $f$ is a function defined on $\mathbb{Z}^{2}_{11}$. Fig. \ref{57} shows the points used to compute $r_{-3}(1)$ as defined in (\ref{223}), i.e., summation along the discrete line $x-3y =1 \mod 11$. Fig. \ref{56} shows the points used to compute $QR_f(1, -3 \mod 2n)$.   }\label{x3}
\end{figure}

\begin{algorithm}
  \caption{ Quantum Radon transform}\label{xradon}
  \begin{algorithmic}[1]
   \Require An $N\times N$ quantum image $\displaystyle{\sum_{x,y\in[N]}} f(x,y)|x\rangle|y\rangle$.
   \Ensure $\displaystyle\sum_{l,j''\in[2N]} QR_f(l,j'')|l\rangle|j''\rangle.$
    \State Prepare the state $\displaystyle{\sum_{x,y\in[N]}} f(x,y)|x\rangle|1\rangle|y\rangle|1\rangle.$
    \State Apply phase shifts conditioned to $|x\rangle,|y\rangle$ to prepare
     \begin{align}
     \displaystyle{\sum_{x,y\in[N]}} g(x,y)|x\rangle|1\rangle|y\rangle|1\rangle,
     \end{align} where $g(x,y)=f(x,y)e^{-2\pi I\frac{x+y}{2N}}$.
    \State Perform quantum Fourier transforms on $|x\rangle$ and $|y\rangle$ respectively to get
    \begin{eqnarray}
     \displaystyle\sum_{i,j\in[N]}\Big{(}\frac{1}{N} \displaystyle\sum_{x,y\in[N]}g(x,y)e^{-2\pi I\frac{ix+jy}{N}}\Big{)}|i\rangle|1\rangle|j\rangle|1\rangle.
     \end{eqnarray}
    \State Perform inverse quantum reversible multiplication on registers $|i'\rangle=|i,1\rangle$ and $|j'\rangle=|j,1\rangle$ to prepare
    \begin{align}\label{cz1}
    &\displaystyle\sum_{i',j''\in[2N]} \Big{(}\frac{1}{2N} \displaystyle\sum_{x',y'\in[2N]} \tilde{f}(x',y')e^{-2\pi I\frac{i'x'+(i'\odot j'') y'}{2N}} \Big{)}\nonumber\\
    &\hspace{6.1cm}|i'\rangle|j''\rangle,
   \end{align}
   where $i'\odot j''=j'$, $\odot$ is the quantum reversible multiplication such that $i'\odot j''= i'j''$ for odd $i'\in[2N]$.
     \State Apply inverse quantum Fourier transform to $|i'\rangle$ to prepare
  \begin{align}
  &\displaystyle\frac{1}{\sqrt{2N}}\sum_{l,j''\in[2N]}  \Big{(}\frac{1}{2N}\displaystyle\sum_{i',t\in[2N]}\displaystyle\sum_{(x',y')\in L^{2N}_{t,j''}}\tilde{f}(x',y')\nonumber\\
   &\hspace{3cm}\times e^{-2\pi I\frac{i' t}{2N}} e^{2\pi I\frac{i' l}{2N}}\Big{)}|l\rangle|j''\rangle\\
   &=\displaystyle\sum_{l,j''\in[2N]} QR_f(l,j'')|l\rangle|j''\rangle.
   \end{align}
 \end{algorithmic}
\end{algorithm}

\subsection{Analysis of Algorithm \ref{xradon}}
%reflected on the lattice $\mathbb{Z}^{2}_{11}$,
The Step 1 of Algorithm \ref{xradon} is adding two extra qubits initially in the state $|1\rangle|1\rangle$. In Step 2, the conditional phase shifts can be implemented in time logarithmic in $N$, by a method similar to the implementation of conditional rotation (cf. Proposition \ref{218}). After applying Step 3, the resulting state can be viewed as the quantum Fourier transform of some $(2N+2)$-qubit state $\displaystyle\sum_{x',y'\in[2N]} g'(x',y')|x'\rangle|y'\rangle$, i.e.,
\begin{eqnarray}
\begin{cases}
&\mathcal{F}_2 \{g'\}(2i+1,2j+1)=\mathcal{F}_2 \{ g \}(i,j), \quad i, j \in [N],\\
&\mathcal{F}_2 \{g'\}(i,j)=0, \quad i, j\in[2N] \text{\ are not both odd}.
\end{cases}
\end{eqnarray}
The explicit expression of function $g'$ can be deduced by inverse Fourier transform as follows:
\begin{align*}
\hspace{0.85cm}g'(l',t')=&\frac{1}{2N}\frac{1}{N}\displaystyle\sum_{x,y,i,j\in[N]}g(x,y)e^{-2 \pi I \frac{ix+jy}{N}}\\
 &\qquad \qquad \qquad \qquad   \qquad \times e^{2 \pi I \frac{(2i+1)l'+(2j+1)t'}{2N}} \nonumber\\
=&\frac{1}{2} g \left(  l' \text{ mod } N,  t' \text{ mod } N \right)e^{2 \pi I \frac{l'+t'}{2N}}\nonumber\\
=&\tilde{f}(l',t'), \qquad \qquad\qquad\qquad\qquad l',t' \in[2N],
\end{align*}
so that the resulting state in Step 3 can be rewritten as
\begin{eqnarray}\label{ccc}
\displaystyle\sum_{i',j'\in[2N]}\Big{(}\frac{1}{2N} \displaystyle\sum_{x',y'\in[2N]}\tilde{f}(x',y') e^{-2\pi I\frac{i'x'+j'y'}{2N}}\Big{)}|i'\rangle|j'\rangle.
\end{eqnarray}
After applying inverse quantum reversible multiplication to $|i'\rangle|j'\rangle$ of (\ref{ccc}) in Step 4, by (\ref{2134}), the resulting state is as in (\ref{cz1}). Now, from the fact that for any $j''\in[2N]$,
\begin{eqnarray*}
\begin{cases}
&i'\odot j''=i' j'', \qquad \qquad \qquad \text{if $i'$ is odd,}\\
&\mathcal{F}_2\{ \tilde{f} \}(i',j'')=0, \qquad \qquad \text{ if $i'$ is even,}
\end{cases}
\end{eqnarray*}
%the fact that $i\odot j''=i j''$ if $i$ is odd and $\mathcal{F}\{ \tilde{f} \}(i,j)=0$ if $i$ is even gives
one gets
\begin{eqnarray*}
\hspace{0.85cm}\mathcal{F}_2 \{ \tilde{f} \}(i',i'\odot j'')=\mathcal{F}_2 \{ \tilde{f} \}(i',i' j''), \qquad i,j\in [2N].
\end{eqnarray*}
Then the resulting state in Step 4 can be rewritten as $\displaystyle\sum_{i',j''\in [2N]} \mathcal{F}_2\{ \tilde{f} \}(i',i' j'')|i'\rangle|j''\rangle$. The final result in Step 5 is by the following proposition:

\begin{prop}\label{cccc1}[Fourier slice property of QRT]
In the notations of Algorithm \ref{xradon}, for any given slope $j''\in[2N]$, the $1$-D Fourier transform of $QR_f(l,j'')$ with respect to $l$ is
\begin{eqnarray}
\mathcal{F}_1 \{ QR_f(\cdot,j'')\} (i')= \mathcal{F}_2 \{ \tilde{f} \}(i', i'j''), \quad  i' \in[2N].
\end{eqnarray}
\end{prop}
\begin{IEEEproof} Let $\mathcal{F}_1^{-1}$ be the inverse transform of $\mathcal{F}_1$. Then,
\setlength{\arraycolsep}{0.0em}
\begin{eqnarray}
&&\mathcal{F}_1^{-1} \{ \mathcal{F}_2 \{ \tilde{f} \} (\cdot, \cdot j'')\}  (l) \nonumber\\
&=&\displaystyle\frac{1}{\sqrt{2N}}  \displaystyle\sum_{i'\in [2N]} \Big{(}\frac{1}{2N}\displaystyle\sum_{x',y'\in [2N]}\tilde{f}(x',y')e^{-2\pi I\frac{i'x'+i'j''y'}{2N}}\Big{)}  e^{2\pi I\frac{i' l}{2N}}  \nonumber\\
&=&\displaystyle\frac{1}{\sqrt{2N}}  \Big{(}\frac{1}{2N}\displaystyle\sum_{i',t\in[2N]}\displaystyle\sum_{(x',y')\in L^{2N}_{t,j''}}\tilde{f}(x',y') e^{-2\pi I\frac{i' t}{2N}} e^{2\pi I\frac{i' l}{2N}} \Big{)} \nonumber\\
&=&\displaystyle\frac{1}{\sqrt{2N}} \displaystyle\sum_{i',t\in[2N]} \Big{(}\frac{1}{2N}\displaystyle\sum_{(x',y')\in L^{2N}_{t,j''}}\tilde{f}(x',y')\Big{)}e^{-2\pi I\frac{i' t}{2N}} e^{2\pi I\frac{i' l}{2N}}   \nonumber\\
&=&\displaystyle\frac{1}{\sqrt{2N}} \displaystyle\sum_{(x',y')\in L^{2N}_{l,j''}}\tilde{f}(x',y')=QR_f(l,j'').
\end{eqnarray}
\end{IEEEproof}

The overall runtime of Algorithm \ref{xradon} is the sum of $O(\text{log} N)$ time required to perform conditional phase shifts \cite{wang2018quantum}, $O(\text{log}^2 N)$ time required to perform quantum Fourier transform and its inverse, and $O(\text{log}^3 N)$ time required to perform the inverse multiplication. Observe that all quantum gates used in Algorithm \ref{xradon} are unitary and invertible. So, an efficient inverse QRT algorithm follows immediately by reversing all circuits in Algorithm \ref{xradon}. In conclusion, we have the following theorem:
\begin{thm}
Given an $N\times N$ quantum image state $\displaystyle{\sum_{x,y\in[N]}} f(x,y)|x\rangle|y\rangle$, then its (inverse) QRT can be performed in time O$(\log^3 N )$.
\end{thm}

%The discretization of Radon transform needs to approximate the line integral. A simple solution is to sample and sum along the integral lines, where

\section{Quantum-mechanical implementation of DRT with the interpolation method}\label{sec3}
The discretization of Radon transform needs to approximate the line integral. A simple solution is to sample and sum along the integral line, where the interpolation method can be employed to estimate the undefined values at non-integer lattice sample points. Several kinds of sample and interpolation methods are available for implementing IDRT \cite{averbuch2001fast,donoho2003digital,kelley1993fast}. Below, we define a simplest kind of IDRT (SIDRT), and give its quantum algorithm, with the aim of showing a general quantum approach to achieving this interpolation-based kind of DRT.
\begin{defn}\label{421}
Let the set of slopes of the basically horizontal lines and basically vertical lines be
\begin{eqnarray}\label{cvq}
&S^{\parallel}:= \{tan \theta_j| \theta_j=\frac{\pi j}{N}, -\frac{N}{4}\leq j<\frac{N}{4}\ \text{and}\ j\in\mathbb{Z} \}, \nonumber\\
&S^{\perp}:= \{\frac{1}{tan \theta_j}| \theta_j=\frac{\pi j}{N}, -\frac{N}{4}< j\leq\frac{N}{4}\ \text{and}\ j\in\mathbb{Z} \},
\end{eqnarray}
respectively, where agree that $\frac{1}{0}=\infty$ and $\frac{1}{\infty}=0$.
Let $\Delta_{ki}$ be the fractional part of $ki$. Then the SIDRT of a function $f$ on $\mathbb{Z}^2_{N}$ is
\begin{eqnarray}\label{vv1}
&\hspace{-1cm}P_{k}(l)=\frac{1}{\sqrt{N}}\displaystyle\sum_{i \in[N]} \Big{[} \sqrt{1-|\Delta_{ki}|^2}f\Big{(}i, l+\lfloor ki \rfloor \text{ mod } N   \Big{)}\nonumber\\
&\hspace{-0.5cm}+\Delta_{ki}f\Big{(}i,l+\lfloor ki \rfloor+1 \text{ mod } N \Big{)}\Big{]}, \quad k\in S^{\parallel}, l \in [N],\\
&\hspace{-1cm}P_{k}(l)=\frac{1}{\sqrt{N}}\displaystyle\sum_{i \in[N]} \Big{[} \sqrt{1-|\Delta_{\frac{i}{k}}|^2}f\Big{(} l+\lfloor \frac{i}{k} \rfloor \text{ mod } N ,i \Big{)}\nonumber\\
&\hspace{-0.5cm}+\Delta_{\frac{i}{k}}f\Big{(}l+\lfloor \frac{i}{k} \rfloor+1\text{ mod } N ,i  \Big{)}\Big{]}, \quad k\in S^{\perp}, l \in [N].\label{vv2}
\end{eqnarray}
\end{defn}
By definition, $P_{k}(l)$ can be viewed as an approximate discrete line integral along the line with interception $l$ and slope $k$.

%Typically, a delicately constructed interpolation kernel guarantees related Fourier slice property hold, such as that in \cite{averbuch2001fast}. %and there is no relevant Fourier slice property or reconstruction formula for the SIDRT.

The SIDRT is proposed for easily achieved in the quantum case. Although it adopts a relatively simple interpolation method, the SIDRT is so useful that enables to detect lines in complicated image, as shown in Figure \ref{xx1}; more details can be found in Section \ref{va1}.

%We devote this scetion to extending the DRT with the interpolation method, based on some recent advances in quantum machine learning and matrix analysis e.g. \cite{kerenidis2017quantum2,shao2018quantum2,wang2018quantum}. Until now, we still fail to find the quantum analogy of the interpolation-based DRT with exponential speedups. However, the polynomial speedup over the classical ones can be archived in some sense (see Corollary \ref{323}). We have given a brief definition of a kind of quantum image representation in Definition \ref{5}. Let us discuss more about the quantum image and presents some

Now, we turn to the quantum implementation of the SIDRT. We first consider approximating the integrals along basically horizontal lines, i.e., $P_k(l)$ as defined in (\ref{vv1}) where $l \in [N]$ and $k \in S^{\parallel}$.

Let $k_\theta:=\tan (\frac{\pi \theta}{N}-\frac{\pi}{4})$. For any $i \in [N]$, $\theta \in[\frac{N}{2}]$, we can prepare a quantum state that contains the location information of lattice points used to compute $P_{k_{\theta}}(l)$, by the following sequence of mappings:
{\small
\begin{align}\label{7}
|i\rangle|\theta&\rangle |0 \rangle |0\rangle \longrightarrow  |i\rangle|\theta\rangle |i k_\theta \rangle |0\rangle \nonumber\\
&\longrightarrow |i\rangle|\theta\rangle \big{|} \lfloor ik_\theta \rfloor \big{\rangle}  |\Delta_{ik_{\theta}}  \rangle  \left (\sqrt{1-|\Delta_{ik_\theta}|^2} |0\rangle + \Delta_{ik_\theta} |1\rangle \right )\nonumber\\
&\longrightarrow |i\rangle|\theta\rangle \big{|} \lfloor ik_\theta \rfloor \big{\rangle}  |0  \rangle  \left ( \sqrt{1-|\Delta_{ik_\theta}|^2} |0\rangle + \Delta_{ik_\theta}|1\rangle \right ).
\end{align}}In (\ref{7}), the first step is by trigonometric function \cite{gilyen2019quantum} and arithmetics \cite{nielsen2000quantum} in computation basis. The second step is to use the decimal part of $ik_\theta$ to perform control rotations (cf. Proposition \ref{218}) on the last qubit. The third step is uncomputing $|\Delta_{ik_{\theta}}  \rangle $.

Combining (\ref{7}) with the addition in the computational basis gives the following lemma:
\begin{lem}\label{2366}
The following `location state' can be prepared in time $O(\emph{polylog} N)$:
\begin{align}\label{cv1}
|\theta\rangle|0\rangle\stackrel{V}{\longrightarrow}&   |\theta\rangle \displaystyle\sum_{i\in[N]} \frac{1}{\sqrt{N}} \Big{(}   \sqrt{1-|\Delta_{ik_\theta}|^2} |i\rangle \big{|} \lfloor ik_\theta \rfloor \big{\rangle}  |0\rangle \nonumber\\
&+ \Delta_{ik_\theta} |i\rangle \big{|} \lfloor ik_\theta \rfloor  +1 \big{\rangle}   |1\rangle  \Big{)}, \quad \theta \in[\frac{N}{2}].
\end{align}
\end{lem}
\emph{Remark:} The state on the right hand side of (\ref{cv1}) is called `location state', since its first two quibts record the locations of points used to compute $P_k(0)$ for $k\in S^{\parallel}$, according to (\ref{vv1}).

%Note the fact that eigenvalues of $R_{U}R_{V_k}$, defined as $R_X:=X(2|0\rangle\langle0|-1)X^{-1}$, are $e^{2I\theta}$
%Therefore, phase estimation allow us to approximate $\displaystyle\sum_{i \in[N]}\left( \Delta_{ki}f(i,\floor{ki})+ \sqrt{1-|\Delta_{ki}|^2}f(i,\floor{ki}+1) \right )$ by using $U, V_k$.

\begin{figure*}[!t]
\centering
\subfloat{\includegraphics[width=2.5in]{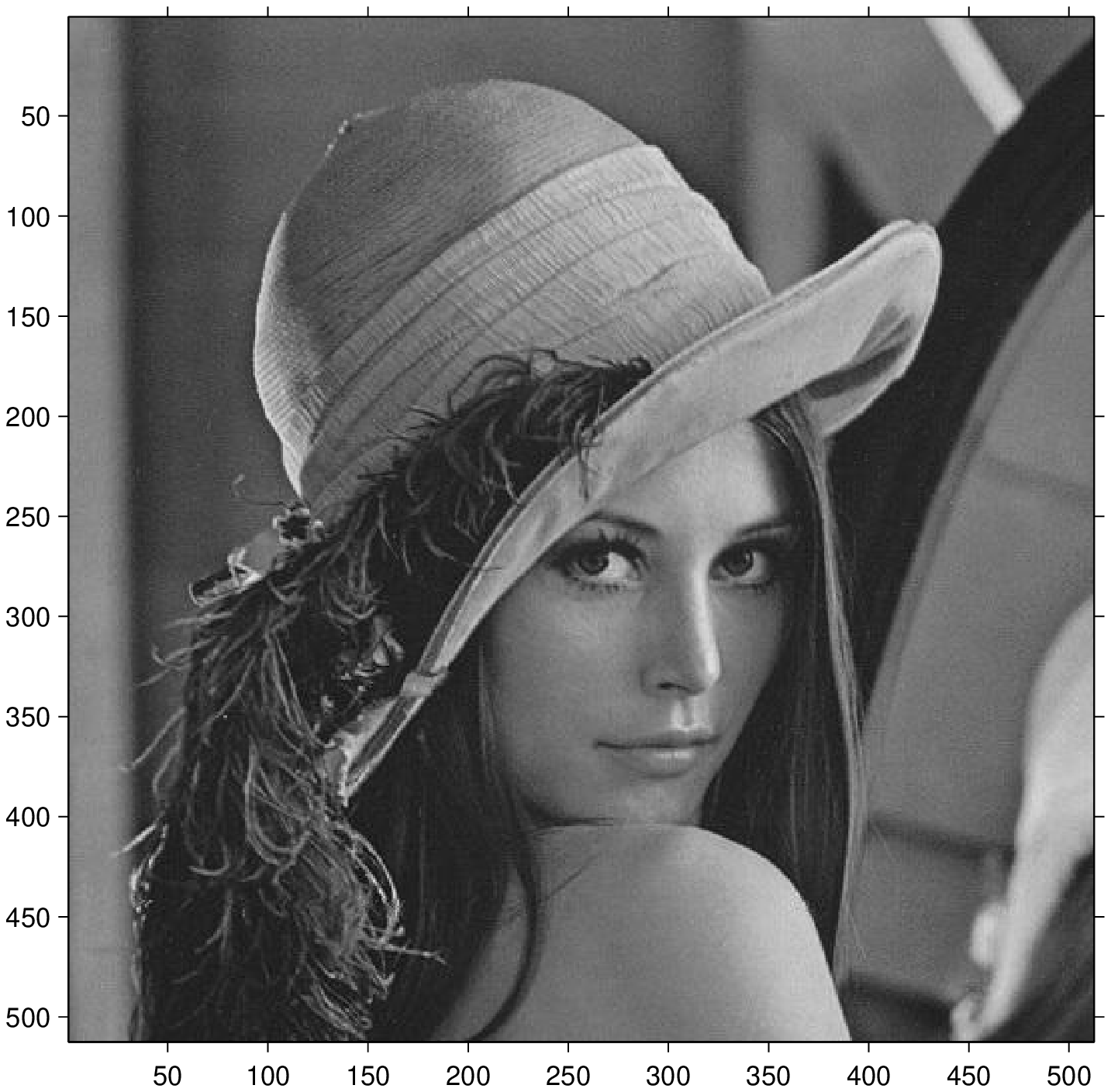}\label{dnq3}}
\subfloat{\includegraphics[width=2.5in]{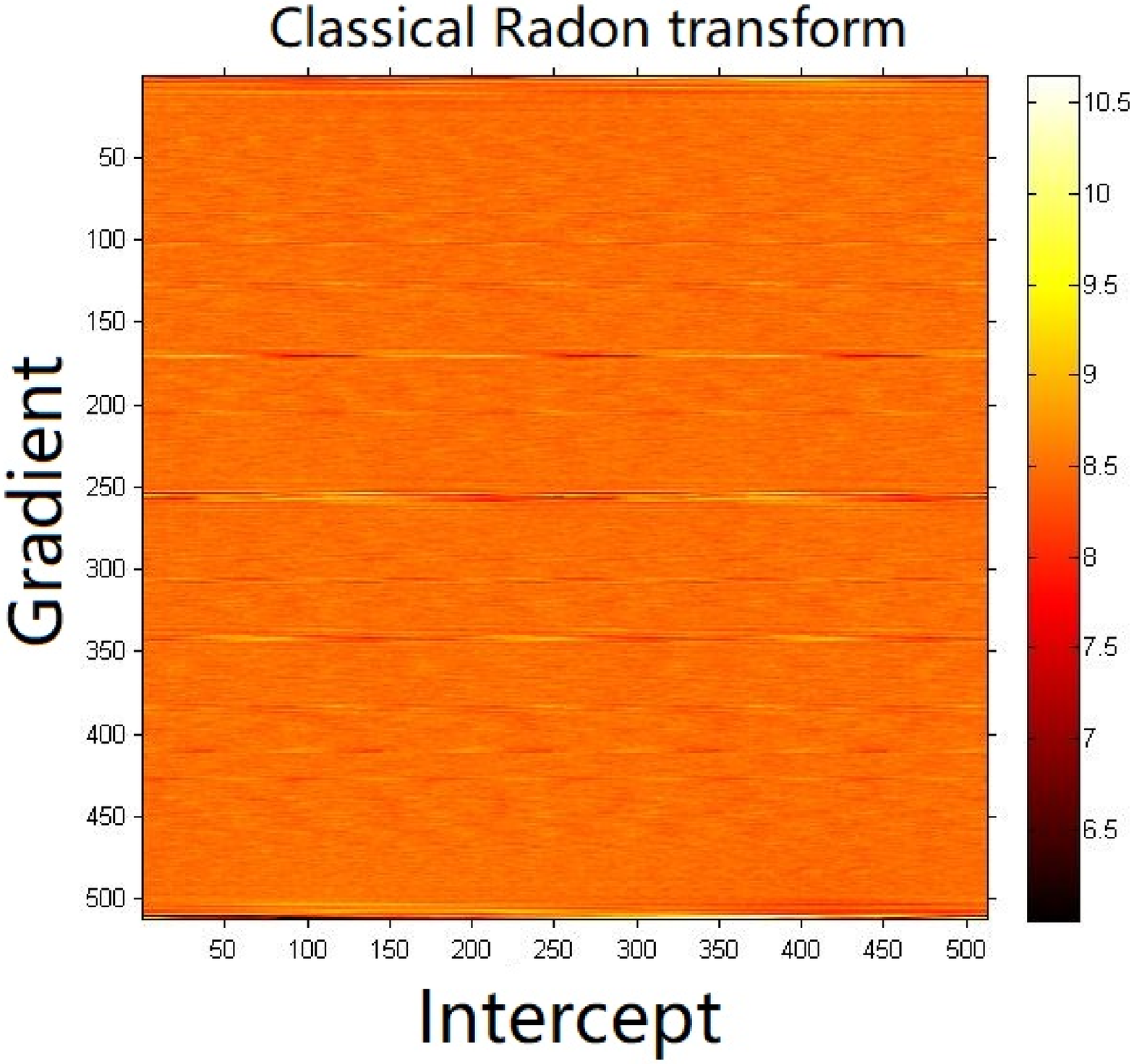}\label{dnq4}}
\subfloat{\includegraphics[width=2.5in]{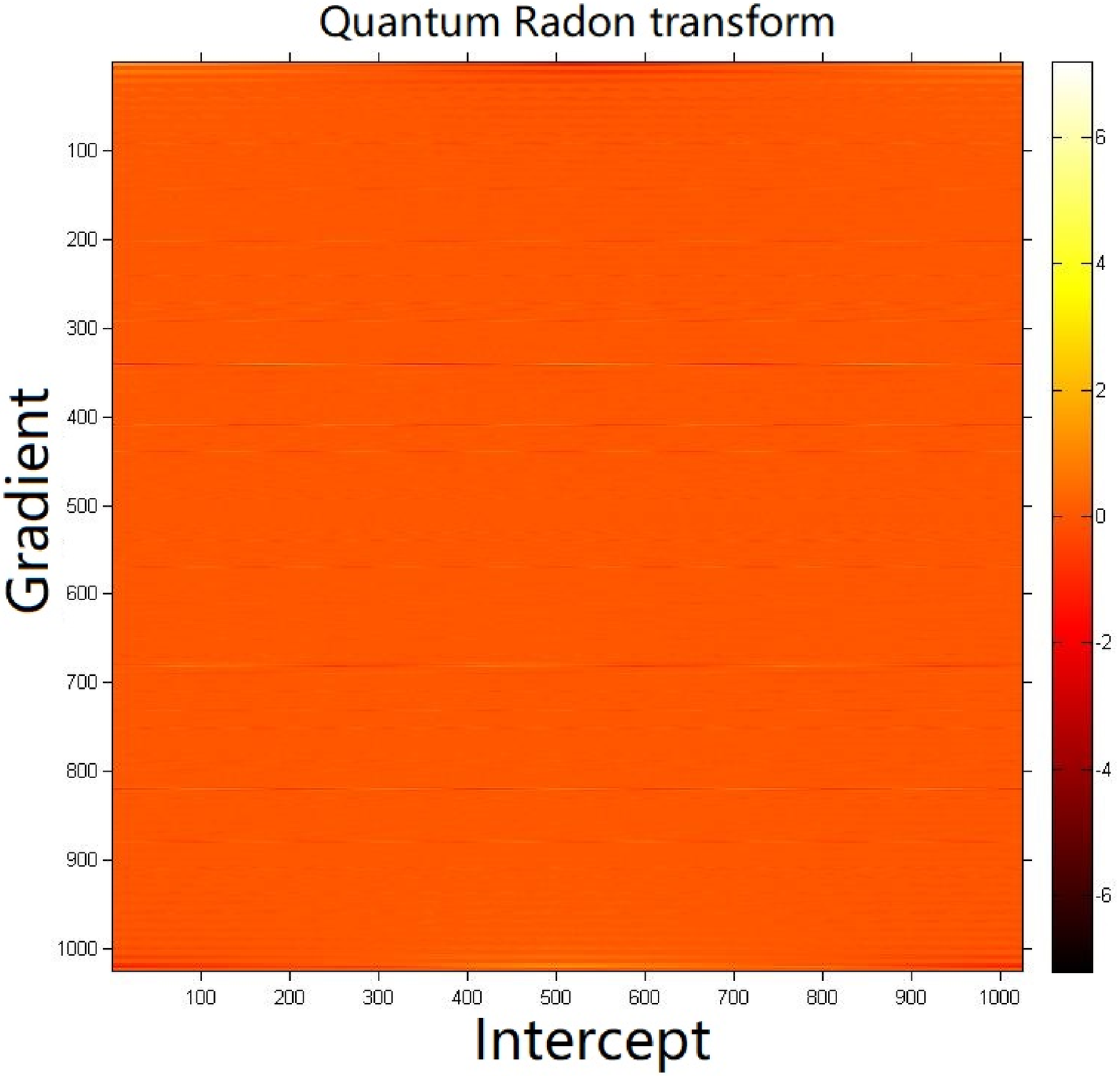}\label{dnq5}}
\caption{\label{11} Left: The `Lena' image. Middle: Classical Radon transform of `Lena'. Right: Quantum Radon transform of 'Lena'.}
\end{figure*}

Now, with the unitary capable of preparing a quantum image $f$ as follows:
\begin{align}\label{zb1}
|0\rangle {\longrightarrow} \displaystyle\sum_{i,j\in[N]} f(i,j)|i\rangle|j\rangle,
\end{align}
one can approximate the SIDRT of $f$ by the following theorem:
\begin{thm}\label{323}
Given the unitary that can prepare a quantum image $\displaystyle\sum_{i,j\in[N]}f(i,j)|i\rangle|j\rangle$ in time O$(T_{in})$, one can approximate $P_k(l)$, the SIDRT of $f$, to precision O$(\epsilon)$ in time O$(\frac{T_{in}}{\epsilon}\emph{polylog} N)$. Namely, there is a quantum algorithm runs in time O$(\frac{T_{in}}{\epsilon}\emph{polylog}N)$ to achieve the mapping
\begin{align}\label{zd1}
|\theta\rangle|l\rangle|0\rangle\rightarrow|\theta\rangle|l\rangle|\tilde{P}_{k_{\theta}}(l)\rangle, \hspace{1cm} \forall l\in[N], \theta\in[N],
\end{align}
where $|\tilde{P}_{k_{\theta}}(l)-P_{k_{\theta}}(l)| \leq  \epsilon$, and $k_\theta:=\tan (\frac{\pi \theta}{N}-\frac{\pi}{4})$ such that $ k_\theta \in S^{\parallel}\cup S^{\perp}$ as defined in (\ref{cvq}).
\end{thm}
\begin{IEEEproof}
We first consider the realization of (\ref{zd1}) for $\theta\in[\frac{N}{2}]$, i.e., the case where $k_\theta\in S^{\parallel}$. Observe that the following `image state' can be prepared in time O($T_{in} \text{polylog} N$) by the mapping
\begin{eqnarray}\label{22}
|\theta\rangle|l\rangle|0\rangle \longrightarrow\displaystyle|\theta\rangle|l\rangle \sum_{i,j\in [N]} f(i,j)|i\rangle|j\rangle (\frac{1}{\sqrt{2}}|0\rangle+\frac{1}{\sqrt{2}}|1\rangle),
\end{eqnarray}
where $\theta\in[\frac{N}{2}]$ and $l\in[N]$. Lemma \ref{2366} allows us to prepare the `location state' by the mapping
\begin{align}\label{23}
|\theta\rangle|l\rangle|0\rangle \longrightarrow |\theta\rangle|l\rangle  \displaystyle\sum_{i\in[N]}& \frac{1}{\sqrt{N}} \Big{(}   \sqrt{1-|\Delta_{ik_\theta}|^2}  |i\rangle \big{|}l+  \lfloor ik_\theta \rfloor \big{\rangle}|0\rangle \nonumber\\
&+ \Delta_{ik_\theta} |i\rangle \big{|} l+ \lfloor ik_\theta \rfloor  +1 \big{\rangle}   |1\rangle  \Big{)}
\end{align}
in time O$(\text{polylog} N)$, where $\theta\in[\frac{N}{2}]$ and $l\in[N]$.

Now, by parallel swap test (Proposition \ref{32}), one can approximate the $\frac{N^2}{2}$ inner products of the `image state' in (\ref{22}) and the `location state' in (\ref{23}) in parallel $\theta,\ l$. Since the result of each inner product is just $\frac{1}{\sqrt{2}} P_{k_\theta}(l)$, one can perform
\begin{eqnarray}
|\theta\rangle|l\rangle|0\rangle\rightarrow|\theta\rangle|l\rangle|p_{k_\theta}(l)\rangle, \hspace{1cm} \forall \theta \in[\frac{N}{2}],\ l \in[N],
\end{eqnarray}
in time O$(\sqrt{2}\frac{T_{in}}{\epsilon}\text{polylog}N)$, where $|p_{k_\theta}(l)-\frac{1}{\sqrt{2}} P_{k_\theta}(l)|\leq \frac{\epsilon}{\sqrt{2}}$. The theorem holds by setting $\tilde{P}_{k_{\theta}}(l)=\sqrt{2}p_{k_\theta}(l)$. An efficient implementation of multiplication by $\sqrt{2}$, a known constant, in the computation basis can be found in \cite{markov2012constant}. The realization of (\ref{zd1}) for $\theta\in[N]\setminus [\frac{N}{2}]$ is similar.
\end{IEEEproof}

%multiplied by a known constant

Given a Real Ket quantum image $f$ and its preparation unitary as in (\ref{zb1}), by Theorem $\ref{323}$, after performing the transform of (\ref{zd1}) on input state $\sum_{\theta,l\in[N]} \frac{1}{N} |\theta \rangle |l\rangle |0\rangle$, one can prepare the SIDRT of $f$ in NEQR encoded form. This NEQR encoded output is already can be used for practical application, such as line detection shown in Section \ref{va1} later. Moreover, if necessary, one can continue to transform the NEQR encoded output into its Real Ket version by Proposition \ref{6}, thus keeping the input and output encoded in the same way.
\subsection{Efficiency Analysis of Theorem \ref{323}}
Below, we discuss what is a reasonable choice of the precision $\epsilon$ in Theorem \ref{323}.

\textbf{Random image.} The term `random image' refers to an image whose each pixel value is sampled from the uniform distribution $\mathbb{U}[0,1]$ independently.

\begin{prop}\label{qwd}
Let $f$ be an $N \times N$ random image, i.e., $f(i,j)\sim\mathbb{U}[0,1]$ for $i$, $j\in [N]$. Then the expectation of the minimal value of the SIDRT of normalized $f$ is no less than $\frac{\sqrt{3}}{2N^{0.5}}$.
\end{prop}
\emph{Remark:} The reason why we consider the normalized rather than the original image here is that the quantum image is a normalized state.

\begin{IEEEproof}
We begin with a probability inequality. Let $a$, $b$ be two discrete random variables whose density function are $P(a=a_i)=p_i, P(b=b_i)=q_i$, where $i\in[N]$, respectively. Then by Cauchy-Schwarz inequality,
\begin{align}\label{vba}
&E(\frac{a}{b})=\sum_{i,j\in [N]}p_iq_j \frac{a_i}{b_j} = \frac{(\sum_{i,j} p_iq_jb_j)(\sum_{i,j} p_iq_j\frac{a_i}{b_j})}{E(b)}\nonumber\\
&\geq \frac{(\sum_{i,j} q_jp_i\sqrt{a_i})^2}{E(b)} = \frac{(\sum_{i}p_i\sqrt{a_i})^2 }{E(b)} = \frac{E^2(\sqrt{a})} {E(b)}.
\end{align}

Let $\mathbb{U}[0,1]$ be the uniform distribution on $[0,1]$, and let random variables $x_k$, $y_k \sim \mathbb{U}[0,1]$ for $k\in[N^2]$. Notice that $f(i,j)\sim\mathbb{U}[0,1]$. Let $P_{k_0}(l_0)$ be the SIDRT of $f$ as in (\ref{vv1}), then for any $k_0$, $l_0 \in[N]$, the expectation of the SIDRT of normalized $f$ at point $(k_0,l_0)$ is
\begin{align}
&\hspace{-0.35cm}E(\frac{P_{k_0}(l_0)}{ \sqrt{\displaystyle\sum_{i,j\in[N]}f^2(i,j) } })=E(\frac{  \frac{1}{\sqrt{N}}\displaystyle\sum_{k\in [N]} \Delta_k x_k+\sqrt{1-|\Delta_k |^2} x_k    }{\sqrt{\displaystyle\sum_{ k\in [N]} x_k^2 + \displaystyle\sum_{ k\in [N^2-N]} y_k^2  }}) \nonumber\\
&\geq E(\frac{  \frac{1}{\sqrt{N}}\displaystyle\sum_{k\in [N]} x_k  }{ \sqrt{\displaystyle\sum_{ k\in [N]} 1 + \displaystyle\sum_{ k\in [N^2-N]} y_k^2  }})
\overset{(\ref{vba})}{\geq} \frac{ E^2(\sqrt{ \frac{1}{\sqrt{N}}\displaystyle\sum_{k\in [N]} x_k  })}{E(  \sqrt{ N + \displaystyle\sum_{ k\in [N^2-N]} y_k^2  } )}\nonumber\\
&\geq \frac{ E^2(\sqrt{ \frac{1}{\sqrt{N}}\displaystyle\sum_{k\in [N]} x_k  })} {  \sqrt{ E(  N + \displaystyle\sum_{ k\in [N^2-N]} y_k^2 )}}\approx \frac{\sqrt{3}}{2 N^{0.5}}  \quad \quad  (N\rightarrow\infty)   ,\label{le1}
\end{align}
where the last inequality follows from that $E^2(\nu)\leq E(\nu^2)$ for arbitrary random variable $\nu$, and the last approximate equality is by the central limit theorem \cite{clt} which states that for any independent and identically distributed (IID) random variables $\nu_0,\nu_1...,\nu_{n-1}$ with mean $\mu$ and variance $\sigma^2$, E$(\sqrt{|\sum_{j\in[n]}\nu_j|})\approx\sqrt{n\mu-\frac{\sigma^2}{\mu}}$ ($n\rightarrow\infty$), and so
$E(\sqrt{\displaystyle\sum_{k\in [N]} x_k  })\approx \sqrt{\frac{N}{2}-\frac{1}{6}}$.
\end{IEEEproof}

The above Proposition implies that a reasonable choice of precision in Theorem \ref{323} is $\epsilon=\Theta(\frac{1}{N^{0.5}})$, e.g., $\frac{1}{100 N^{0.5}}$, for producing a good approximation to SIDRT. In this case, when compared with the classical SIDRT whose running time is $\Omega(N^{3})$, quantum SIDRT achieves a polynomial speedup by Theorem \ref{323}, because the quantum image preparation time $T_{in}\leq O(N^2)$ in the worst case, thus $\frac{T_{in}}{\epsilon}\leq O(N^{2.5}) < N^{3}$.

%In comparison with the classical time O$(N^{3})$ required to straightforwardly compute the SIDR, we can witness a quantum polynomial speedup, if the time complexity for image state preparation $T_{in}$ divided by the desired precision $\epsilon$ is strictly less than O$(N^{3})$. Indeed, the choice of precision is dependent on the practical issues, which could be set at a level of O($\frac{1}{\sqrt{N}}$) in some cases. To see this, note that as the normalized image intensity $f(i,j)$ should satisfy $\sum_{i,j} (f(i,j))^2=1$. We simply assume that each value of $f$ has an expected level of $\frac{1}{N}$, and thus the maximum value of its SIDRT computed by Definition \ref{421} could reach $\frac{1}{\sqrt{N}}$. This implies the desired precision of SIDRT could be set at a level such as $\frac{1}{100\sqrt{N}}$, to only ensure that we will obtain sufficiently good approximations to those high values which could reach $\frac{1}{\sqrt{N}}$, since only the information of these high values is concerned in some practical tasks like line detection.
\section{Quantum application}\label{sec4}
We present two potential applications---quantum image denoising using QRT and quantum line detection using SIDRT.
\subsection{Denoising using QRT}\label{sec4.1}
\begin{figure*}[!t]
\hspace{-0.3cm}\subfloat{\includegraphics[width=7.3in]{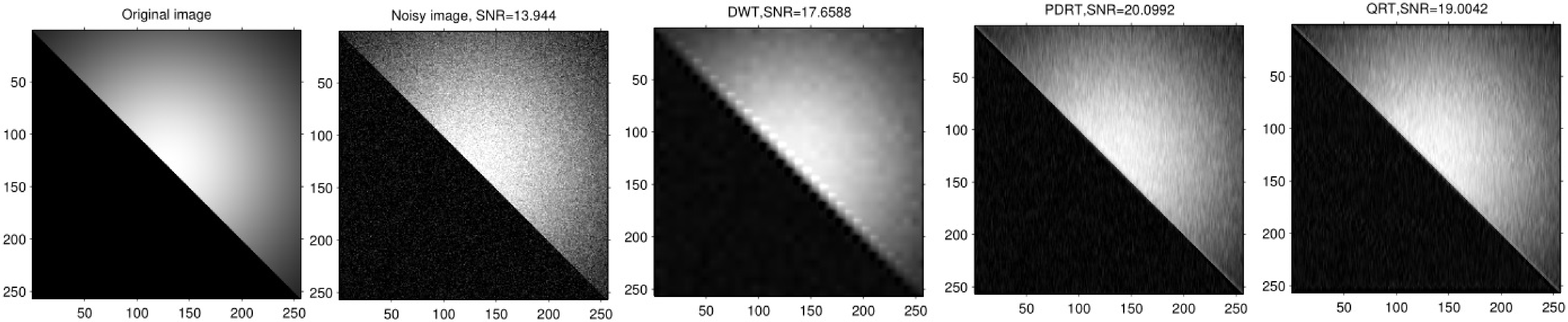}\label{dnqi}}
%\subfloat{\includegraphics[width=1.75in]{dnq2.eps}\label{dnq2}}
%\subfloat{\includegraphics[width=1.75in]{dnq3.eps}\label{dnq3}}
%\subfloat{\includegraphics[width=1.75in]{dnq4.eps}\label{dnq4}}
%\subfloat{\includegraphics[width=1.75in]{dnq5.eps}\label{dnq5}}
\caption{Comparison of denoising half-plane truncated Gaussian function $f(x_1,x_2)=1_{\{x_1>x_2\}}e^{(x_1-128)^2-(x_2-128)^2}$. By denoising using QRT, the SNR increases 5.0602. The QRT, with a comparable performance to PDRT, is better than DWT at handling the image which is piecewise smooth with singularities along a straight line (here is the diagonal line).}\label{21}
\hfil
\centering
\subfloat{\includegraphics[width=2.5in]{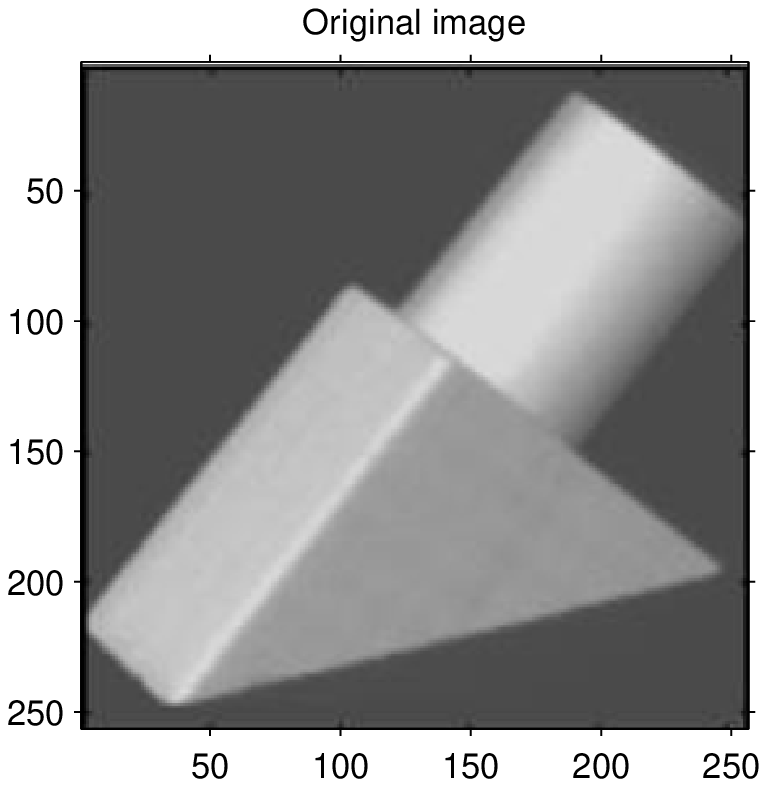}\label{foi}}
\subfloat{\includegraphics[width=2.5in]{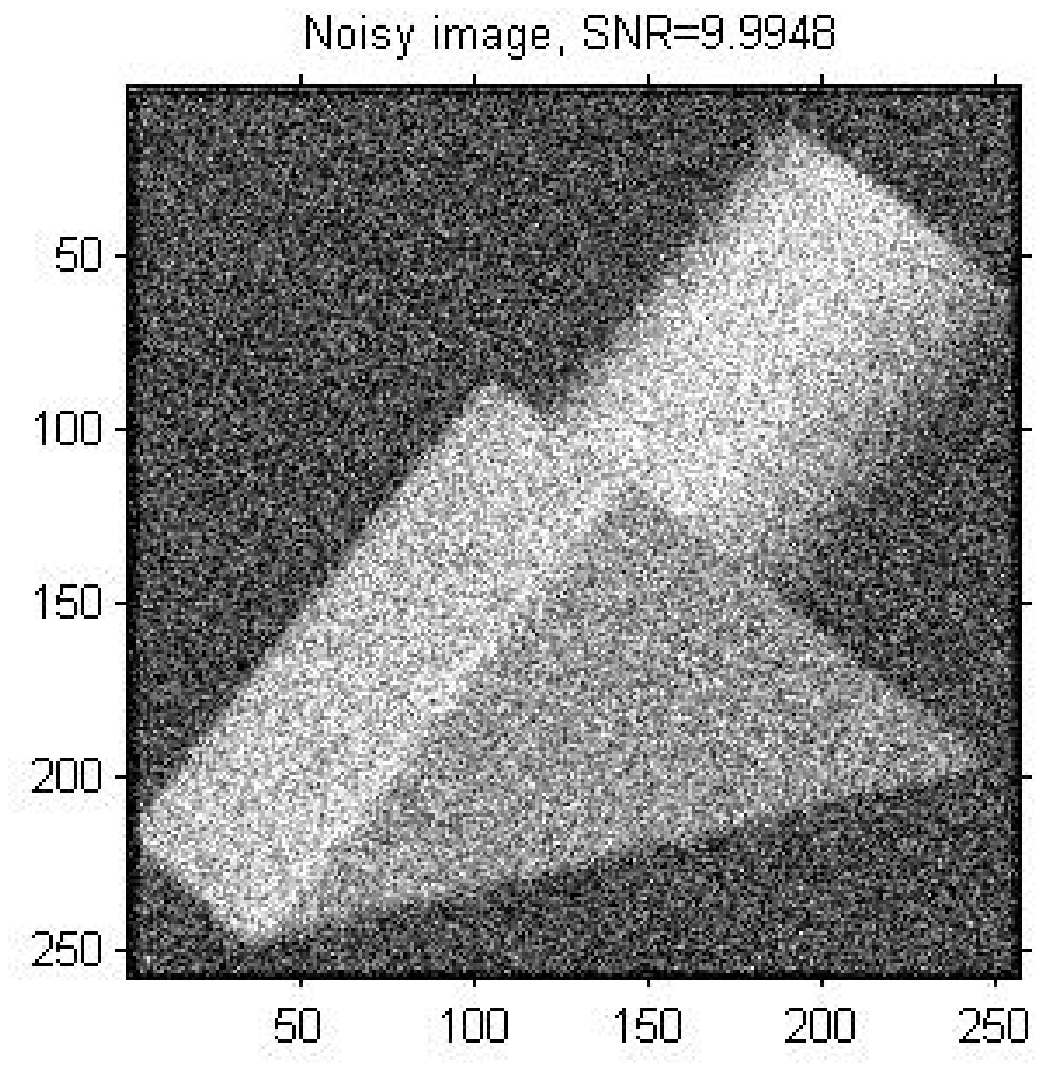}\label{fni}}
\hfil
\subfloat{\includegraphics[width=2.5in]{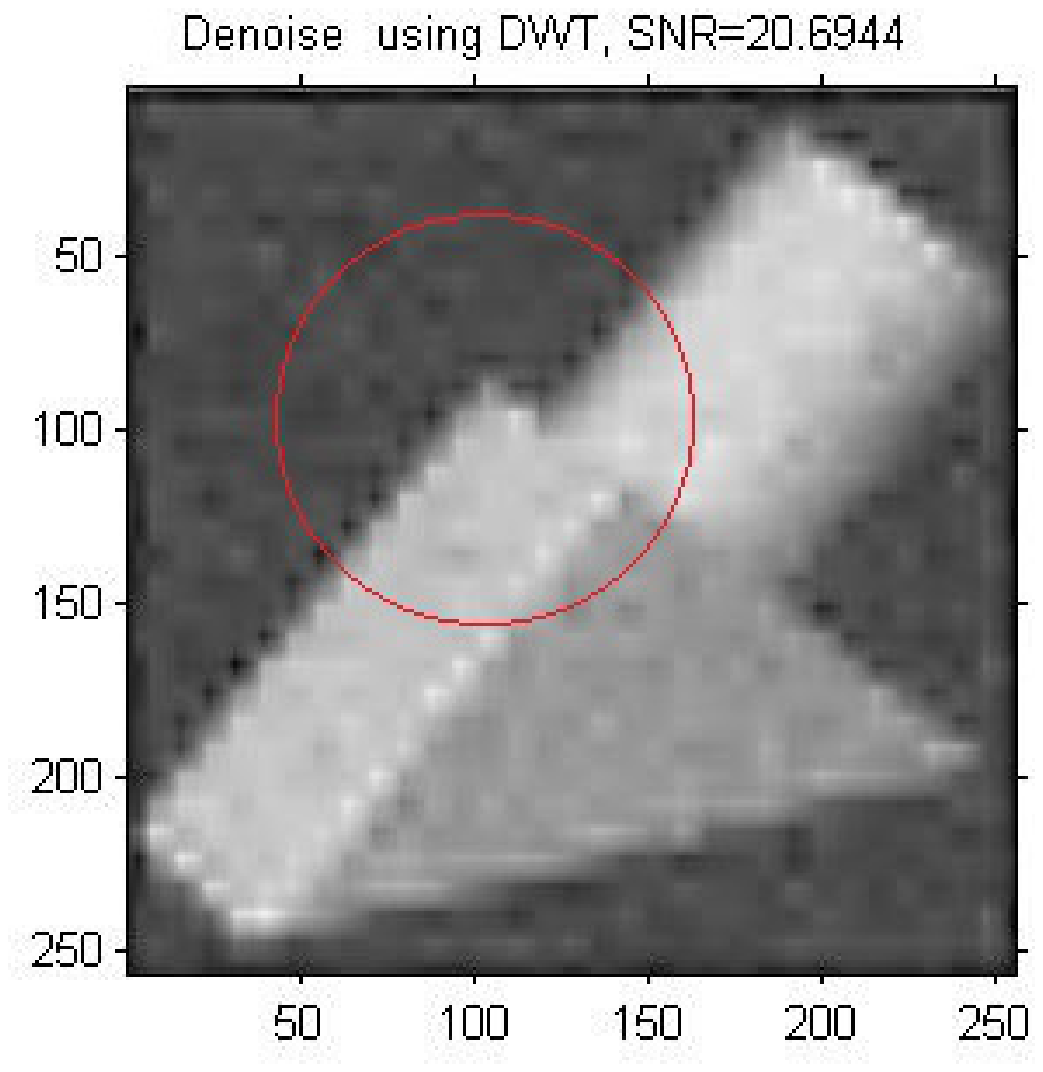}\label{fd2}}
\subfloat{\includegraphics[width=2.5in]{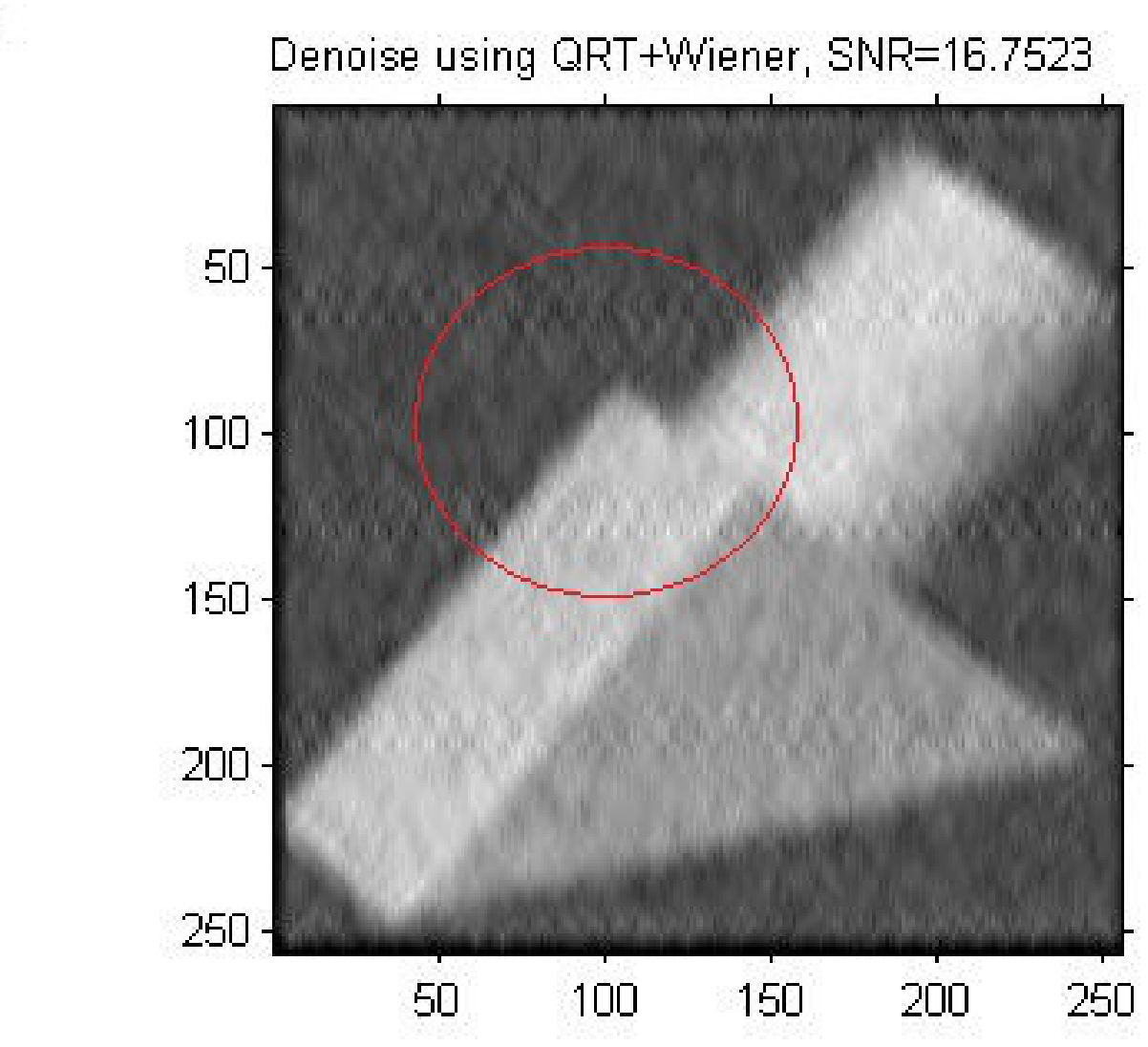}\label{fq2}}
\subfloat{\includegraphics[width=2.5in]{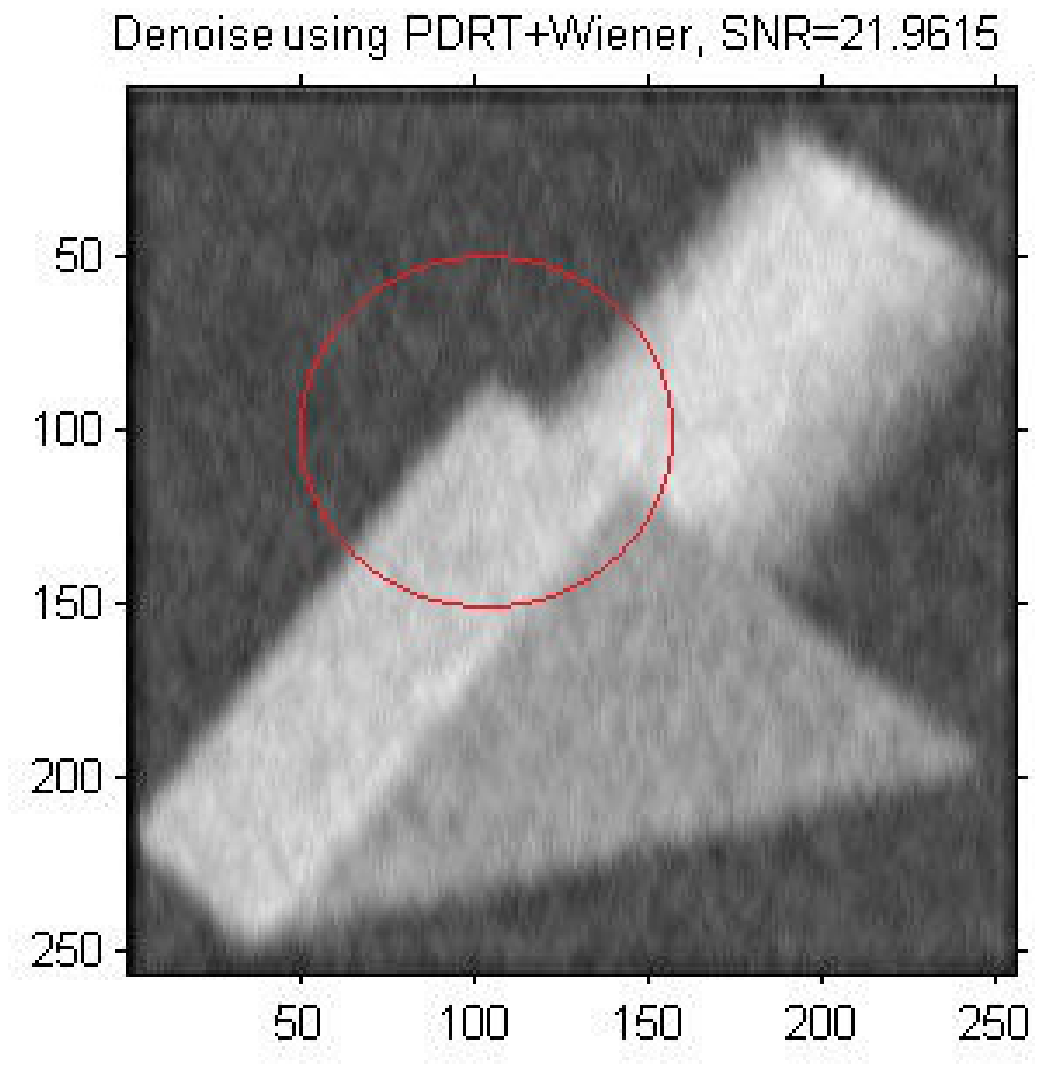}\label{fp2}}
\caption{In the area marked by the red circles, it can be seen that the QRT, similar to PDRT, enables to restore the straight edges of geometric solids more clearly than DWT, even in a lower total denoising performance.}
\label{fig:image10}
\end{figure*}

We replicate the denoising experiments in \cite{do2000image}, which are specially designed for testing PDRT. The experiment results shown in Fig. \ref{21}-\ref{fig:image10} suggest that our QRT is of the good denoising capability as in the classical PDRT. Below, we give an efficient quantum image denoising algorithm using QRT.

\begin{figure*}[!t]
\centering
\subfloat[Case I]{\includegraphics[width=2.5in]{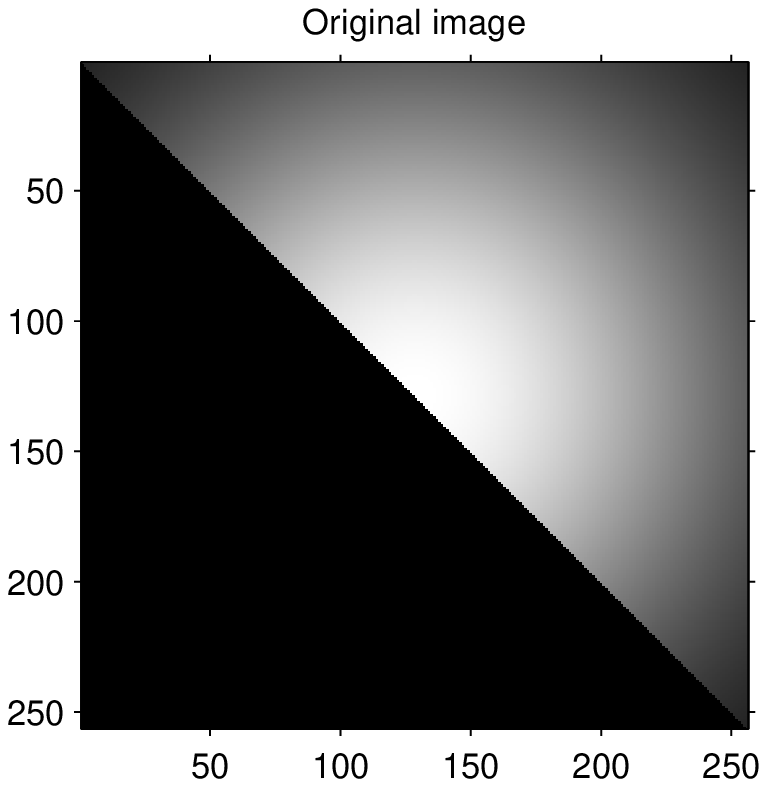}\label{dnqi}}
\subfloat[Case I]{\includegraphics[width=2.5in]{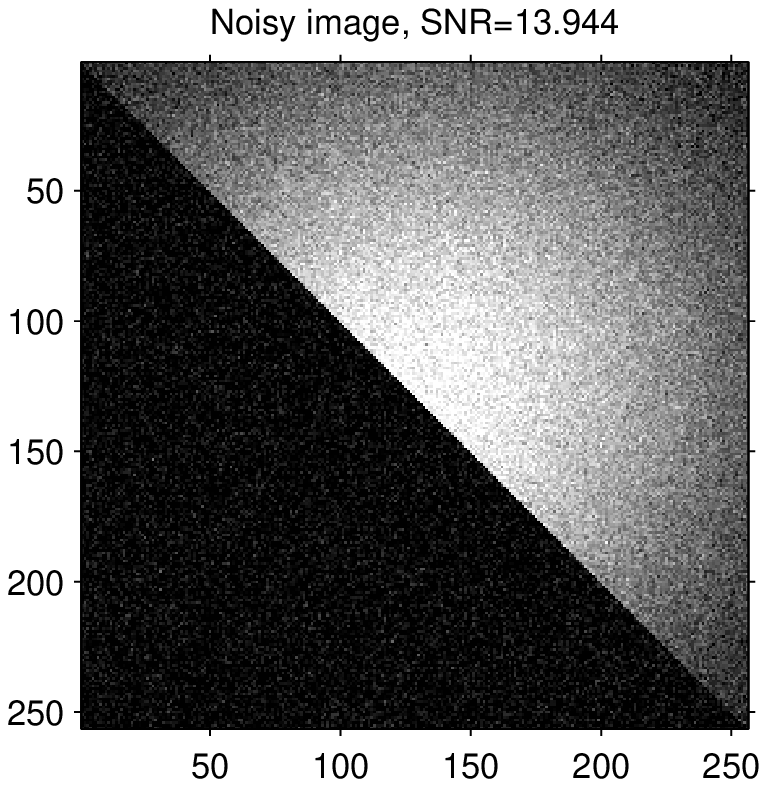}\label{dnq2}}
\hfil
\subfloat[Case I]{\includegraphics[width=2.5in]{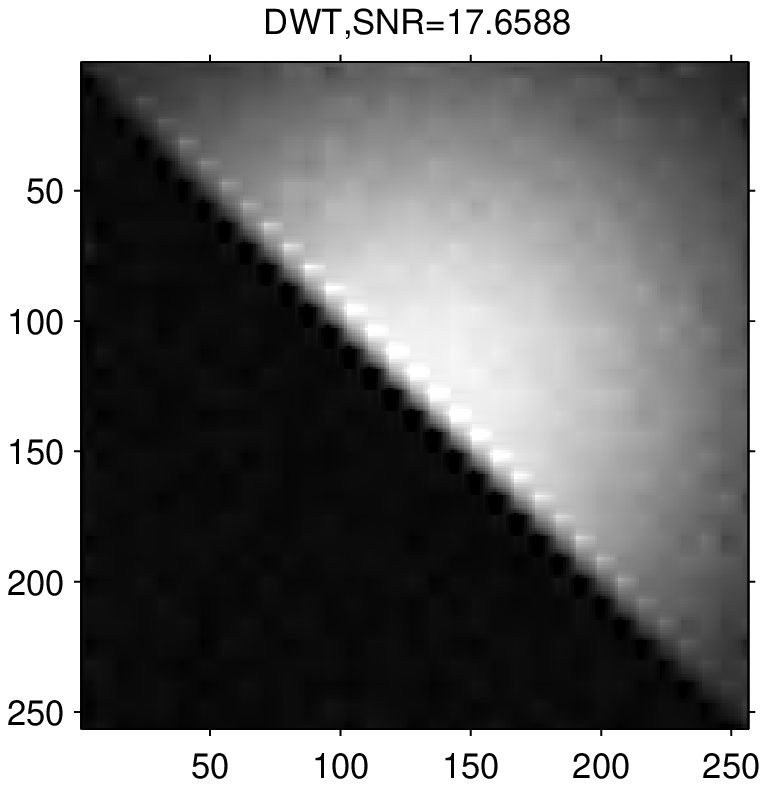}\label{dnq3}}
\subfloat[Case I]{\includegraphics[width=2.5in]{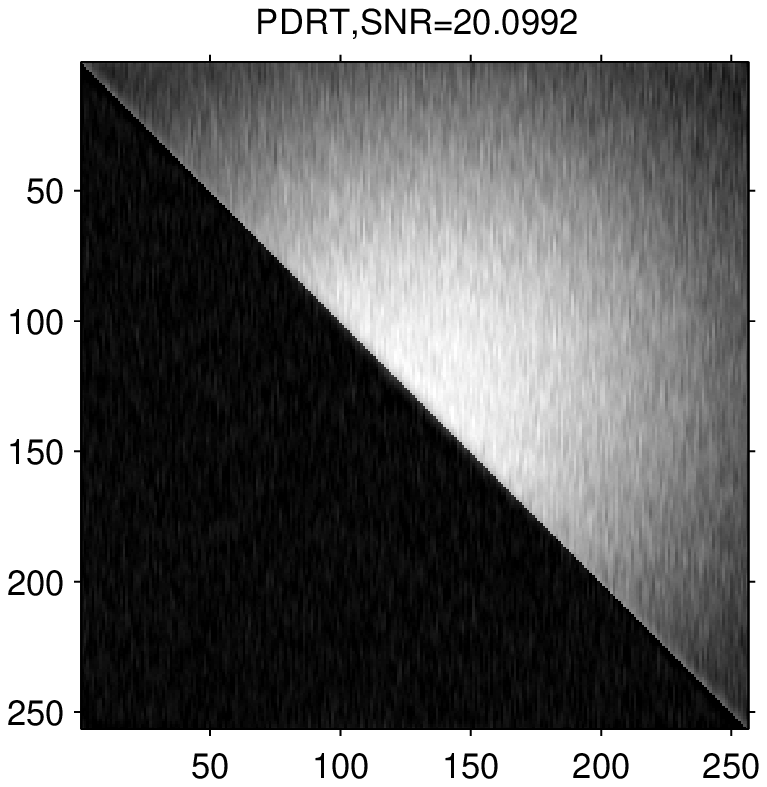}\label{dnq4}}
\subfloat[Case I]{\includegraphics[width=2.5in]{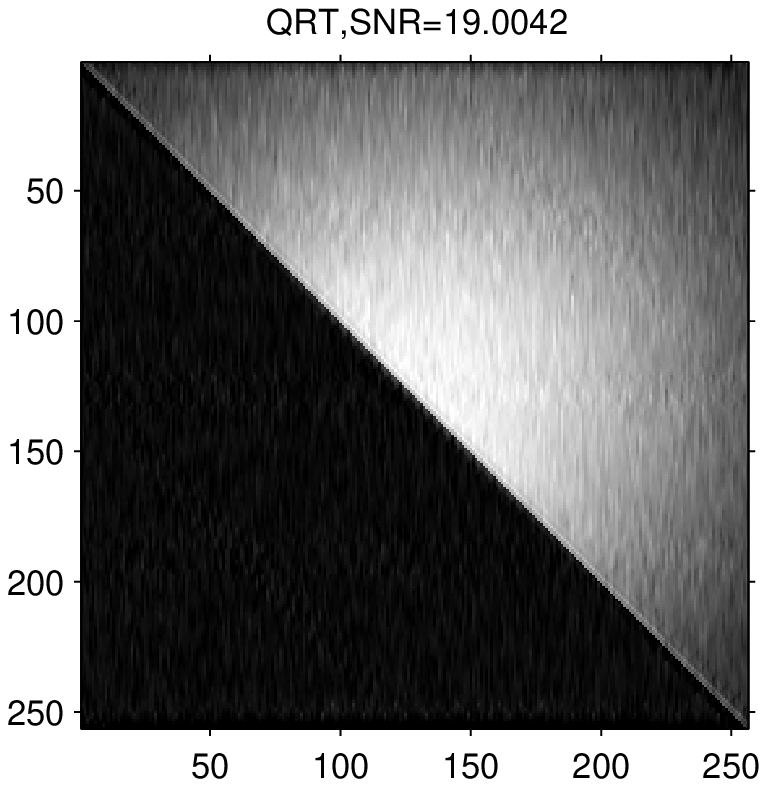}\label{dnq5}}
\caption{Comparison of denoising half-plane truncated Gaussian function $f(x_1,x_2)=1_{\{x_1>x_2\}}e^{(x_1-128)^2-(x_2-128)^2}$. By (\ref{snr}), a higher singal-noise raito (SNR) indicates a better performance of denoising. So, QRT, with a comparable performance to PDRT, is better than DWT at handling the image which is piecewise smooth with singularities along a straight line (here is the diagonal line).}
\label{21}
\end{figure*}

%In this experiment, the chosen threshold is $1$, the wavelet is Daubechies-$4$ wavelet, and reconstruction level is $3$.

%\begin{figure}
%\begin{minipage}[t]{0.4\linewidth}
%\centering
%\includegraphics[width=1.5in]{dnq1.eps}\caption{Original image.}\label{s1}
%\end{minipage}
%\hspace{5mm}
%\begin{minipage}[t]{0.4\linewidth}
%\centering
%\includegraphics[width=1.5in]{dnq2.eps}\caption{Noisy image. SNR = $13.944$.}\label{s2}
%\end{minipage}
%\begin{minipage}[t]{0.4\linewidth}
%\centering
%\includegraphics[width=1.5in]{dnq3.eps}\caption{Denoise using 2D-DWT. SNR = $17.6588$.}
%\end{minipage}
%\hspace{5mm}
%\begin{minipage}[t]{0.4\linewidth}
%\centering
%\includegraphics[width=1.5in]{dnq4.eps}\caption{Denoise using PDRT. SNR = $20.0922$.}
%\end{minipage}

%\includegraphics[width=1.5in]{dnq5.eps}\caption{Denoise using QRT. SNR = $19.0042$.}

%\caption{\label{21} Comparison of denoising on half-plane truncated Gaussian function $f(x_1,x_2)=1_{\{x_1>x_2\}}e^{(x_1-128)^2-(x_2-128)^2}$. In all above denoising methods, the thresholds are set at $T=1$, the wavelets are Daubechies4 wavelet, and reconstruction levels are $3$. The higher Singal-Noise raito (SNR) indicates the better performance of denoising. We can see that the QRT yields a performance comparable with the classical PDRT, which is better than that of DWT, where the test image is piecewise smooth with singularities along a straight line,}
%\end{figure}

\begin{itemize}[\IEEEsetlabelwidth{Step}]\label{qdm}
\item[Step 1.] Apply the QRT algorithm on the input $N\times N$ quantum image $f$ to perform
\begin{align}\label{cc}
\displaystyle{\sum_{x,y\in[N]}} f(x,y)|x\rangle|y\rangle \longrightarrow \displaystyle\sum_{l,k\in[2N]} QR_f(l,k)|l\rangle|k\rangle.
\end{align}

\item[Step 2.] Perform Hadamard gate (namely, $2\times2$ Haar transform as shown in Fig. \ref{xfig:2}) on the least significant qubit (LSB) of $|l\rangle$ in (\ref{cc}) to produce a state
\end{itemize}
\vspace{-0.5cm}
\begin{itemize}
\item[]
\begin{align}\label{gc}
\frac{1}{\sqrt{2}} &\displaystyle\sum_{l'\in[N],k\in[2N]} \big{(}QR_f(2l',k)+QR_f(2l'+1,k)\big{)}|l'0\rangle|k\rangle\nonumber\\
&+\big{(}QR_f(2l',k)-QR_f(2l'+1,k)\big{)}|l'1\rangle|k\rangle.
\end{align}
\end{itemize}

\begin{itemize}[\IEEEsetlabelwidth{Step}]
\item[Step 3.] Measure the LSB of the first register. If the outcome is $|0\rangle$, then apply the inverse Hadamard gate to the measured qubit, and then execute the inverse QRT algorithm.
The resulting state is (up to a normalization factor):
\begin{align*}
\hspace{-1.2cm} QR^{-1}\Big{(}&\displaystyle\sum_{\substack{ l'\in[N], s\in[2] \\ k\in[2N]}}\frac{1}{2} \Big{(}QR_f(2l',k)
+QR_f(2l'+1,k)\Big{)}|l's\rangle|k\rangle\Big{)}
\end{align*}
\end{itemize}

The above is a quantum analogue of PDRT denoising method introduced in Section \ref{41}. Notice that the Hadamard gate in Step 2 plays the role of Haar wavelet, and measuring LSB to obtain $|0\rangle$ in Step 3 has the effect of making all wavelet coefficients zero. So, this quantum QRT denoising method indeed simulates denoising using QRT$+$Haar wavelet and threshold $\infty$.

\begin{figure}
\centering
\includegraphics[scale=0.5]{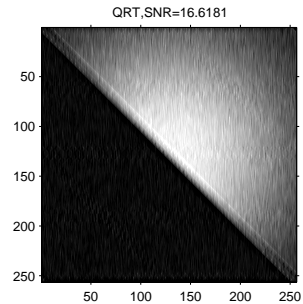}
\caption{Denoise using QRT$+$Haar wavelet with setting threshold value $\infty$ (i.e., make all wavelet coefficients zeros). The original and noisy image are as shown in Fig. \ref{21}. By denoising, the SNR increases 2.674. The ratio of ($L^2$-norm of) wavelet coefficients to filter coefficients is 0.0422, which means the probability of success of the quantum denoising algorithm on Section \ref{sec4.1} is higher than 95$\%$ for this example.}\label{s3}     % Give a unique label
\end{figure}

To evaluate the efficiency of this denoising method, we consider the success probability (namely, the probability of measuring $|0\rangle$ in Step 3). Since $QR_f(l',k)=0$ holds for any even $k\in[2N]$, the success probability is thus reduced to:
\begin{align}\label{42}
\frac{1}{2} \displaystyle\sum_{l',k\in[N]} |QR_f(2l',2k+1)+QR_f(2l'+1,2k+1)|^2.
\end{align}
By the following theoretical analysis (cf. Proposition \ref{b1}) and numerical simulation (cf. Fig. \ref{s3}), this probability of success is quite high, particularly, it has no tendency to be tiny as the image size $N$ increases.
\begin{prop}\label{b1}
Let $f$ be an $N \times N$ random image, i.e., $f(i,j)\sim\mathbb{U}[0,1]$ for $i$, $j\in [N]$. Given the noisy image $h(i,j)=f(i,j)+\epsilon e_{ij}$, $i,j\in[N]$, where $e_{ij}$ is the noise sampled from the normal distribution $V(0,\sigma^2)$ independently, and $\epsilon\in\mathbb{R}$ is noise level, then the QRT denoising method can be implemented with an average probability of success $p>1/2$.
\end{prop}

\begin{IEEEproof}
By Definition \ref{1021}, for any fixed $l,k \in [2N]$,
\begin{align}\label{z1}
QR_h(l,k)-QR_f(l,k)=\frac{1}{2\sqrt{2N}}\sum_{(i,j)\in L^{2N}_{l,k}} \epsilon e_{ij}.
\end{align}
Notice that for two independent random variables $x\sim V(u_1,\sigma_1^2)$, $y\sim V(u_2,\sigma_2^2)$, it holds that $x\pm y \sim V(u_1\pm u_2,\sigma_1^2+\sigma_2^2)$. So, the distribution of the right-hand side of (\ref{z1}) is $\epsilon V(0,\sigma^2/4)$. Let $E_{f,e}(QR_h)$ denote the expectation of $QR_h$ with respect to the random variables $f,e$. Then, for any $l,k \in [2N]$,
\begin{align}
&E_{f,e}(|QR_h(l,k)-QR_h(l+1,k)|^2)\nonumber\\
=& E_{f}(|QR_f(l,k)-QR_f(l+1,k)|^2)\nonumber\\
&\hspace{2cm}+E_{e}(|\frac{\epsilon}{2\sqrt{2N}}\sum_{s\in[2N]}e_s-\sum_{s\in[2N]}e_s'|^2)\label{fe1}\\
=& E_{f}(|QR_f(l,k)-QR_f(l+1,k)|^2)+ \epsilon^2\sigma^2/2,\label{kk1}
\end{align}
where $e_s,\ e'_s\sim V(0,\sigma^2)$ are both the short hands of IID noises $e_{ij}$ in (\ref{z1}). (\ref{fe1}) is by combining (\ref{z1}) and the fact that the mean of random variables $e_s,\ e'_s$ is $0$.
Similarly,
\begin{align}\label{kk2}
&E_{f,e}(|QR_h(l,k)+QR_h(l+1,k)|^2)\nonumber\\
&= E_{f}(|QR_f(l,k)+QR_f(l+1,k)|^2)+ \epsilon^2\sigma^2/2.
\end{align}
Now, we denote
\begin{eqnarray*}
&\Delta_{+}=\displaystyle\sum_{l,k\in[N]} |QR_h(2l,2k+1)+QR_h(2l+1,2k+1)|^2\\
&\Delta_{-}=\displaystyle\sum_{l,k\in[N]}|QR_h(2l,2k+1)-QR_h(2l+1,2k+1)|^2\\
&\Delta'_{+}=\displaystyle\sum_{l,k\in[N]} |QR_f(2l,2k+1)+QR_f(2l+1,2k+1)|^2\\
&\Delta'_{-}=\displaystyle\sum_{l,k\in[N]} |QR_f(2l,2k+1)-QR_f(2l+1,2k+1)|^2.
\end{eqnarray*}
From (\ref{42}), our aim is to lower bound the following kind of average probability of success:
\begin{eqnarray}\label{213}
&P_{\text{success}}= \frac{E_{f,e}( \Delta_+)}{ E_{f,e} (\Delta_{+} +\Delta_{-})}\nonumber\\
&\xlongequal{(\ref{kk1}),\ (\ref{kk2})}\frac{E_{f}(\Delta'_+)+N^2\epsilon^2\sigma^2/2}{ E_{f}(2\sum_{i,j\in[N]}|f(i,j)|^2)+N^2\epsilon^2\sigma^2},
\end{eqnarray}
where the denominator in (\ref{213}) is by the following relations:
\begin{align}\label{kk5}
\sum_{l,k\in[N]} \Delta'_+ + \Delta'_- =\sum_{l,k\in[2N]} 2|QR_f(l,k)|^2=2\sum_{i,j\in[N]}|f(i,j)|^2
\end{align}
%\begin{widetext}
%\begin{eqnarray}\label{213}
%P_{\text{success}}&=E_{f,e}\left( \frac{\displaystyle\sum_{l,k\in[N]} |QR_h(2l,2k+1)+QR_h(2l+1,2k+1)|^2}{\displaystyle\sum_{l,k\in[N]}|QR_h(2l,2k+1)+QR_h(2l+1,2k+1)|^2+|QR_h(2l,2k+1)-QR_h(2l+1,2k+1)|^2}\right)\nonumber\\
%&\geq\frac{E_{f}(\displaystyle\sum_{l\in[N],k\in[N]}|QR_f(2l,2k+1)+QR_f(2l+1,2k+1)|^2)+N^2\epsilon^2\sigma^2/2}{ E_{f}(|f|^2)+N^2\epsilon^2\sigma^2}
%\end{eqnarray}
%\end{widetext}

Notice that if $k$ is even, then $QR_f(l,k)\equiv0$. We first consider estimating $E_{f}(|QR_f(l,k)+QR_f(l+1,k)|^2)$ and $E_{f}(|QR_f(l,k)-QR_f(l+1,k)|^2)$ for any odd $k\in[2N]$. Denote intervals
\begin{eqnarray}
P_1:=[0,N)\times[0,N) \bigcup [N,2N)\times[N,2N),\nonumber\\
P_2:=[0,N)\times[N,2N) \bigcup [N,2N)\times[0,N),
\end{eqnarray}
and denote the number of points in $L^{2N}_{l,k}\bigcap P_1$, $L^{2N}_{l,k}\bigcap P_2$ by
\begin{align}
C_1(L^{2N}_{l,k})=\text{card}( \{(x,y)\in L^{2N}_{l,k}|(x,y)\in P_1\}),\nonumber\\
C_2(L^{2N}_{l,k})=\text{card}( \{(x,y)\in L^{2N}_{l,k}|(x,y)\in P_2\}),
\end{align}
respectively. By geometry, it can be verified that
\begin{align}
&C_1(L^{2N}_{l,k})+C_2(L^{2N}_{l,k})=2N, \quad \forall l,k \in[2N], \label{25}\\
&|C_1(L^{2N}_{l,k})-C_1(L^{2N}_{l+1,k})|=2,\quad \forall \textnormal{ odd } k\in[2N].\label{24}
\end{align}

By (\ref{021}), $QR_f(l,k)$ is a sum of values $\tilde{f}(i,j)$ over line $L^{2N}_{l,k}$ on $\mathbb{Z}^2_{2N}$, where each $\tilde{f}(i,j)$ takes values from uniform distribution on $[0,1]$ for $(i,j) \in P_1$, and each $\tilde{f}(i,j)$ takes values from uniform distribution on $[-1,0]$ for $(i,j) \in P_2$. Combining (\ref{25}) and (\ref{24}) gives that for any odd $k$,
\begin{eqnarray}\label{vz2}
&&E_{f}(|QR_f(l,k)-QR_f(l+1,k)|^2)\nonumber\\
&=&\frac{1}{8N}E(\big{|}\sum\limits_{C_1(l)+C_2(l+1)}\nu-\displaystyle\sum_{C_2(l)+C_1(l+1)}\nu\big{|}^2)\nonumber\\
&=&\frac{1}{8N}E(\big{|}\displaystyle\sum_{2N+2}\nu-\displaystyle\sum_{2N-2}\nu\big{|}^2),
\end{eqnarray}
where $C_1(l)$ is the short hand of $C_1(L^{2N}_{l,k})$, and $\sum_{k}\nu$ denotes the summation of $k$ numbers of independent random variables $\nu\sim\mathbb{U}[0,1]$.
%\begin{align*}
%&E_{f}(|QR_f(l,k)+QR_f(l+1,k)|^2)\\
%&=\frac{1}{8N}E(\big{|}\displaystyle\sum_{2N-C_2(l)+C_1(l+1)}\chi-\displaystyle\sum_{2N+C_2(l)-C_1(l+1)}\chi\big{|}^2)
%\end{align*}
Notice that for IID random variables $\nu_1,\nu_2\sim\mathbb{U}[0,1]$, we have $E(|\nu_1|^2)=|E(\nu_1)|^2+D(\nu_1)$, where $D$ is the variance, $D(\nu_1\pm\nu_2)=2 D(\nu_1)$, and $E(\nu_1\pm\nu_2)=E(\nu_1)\pm E(\nu_1)$. So,
\begin{align}\label{vz3}
\hspace{-0.5cm}E(\big{|}\displaystyle\sum_{2N+2}\nu-\displaystyle\sum_{2N-2}\nu&\big{|}^2)=D(\sum_{2N+2}\nu-\displaystyle\sum_{2N-2}\nu)\nonumber\\
&+E(\sum_{2N+2}\nu-\displaystyle\sum_{2N-2}\nu)=\frac{N}{3}+2.
\end{align}
Combining (\ref{vz3}) and (\ref{vz2}) gives
\begin{align}\label{vz7}
&E_{f}(\displaystyle\sum_{l\in[N],k\in[N]}|QR_f(2l,2k+1)-QR_f(2l+1,2k+1)|^2)\nonumber\\
&=N^2\times\frac{1}{8N}\times(\frac{N}{3}+2)=\frac{N^2}{24}+\frac{N}{4}
\end{align}

Then by (\ref{213}), (\ref{kk5}), (\ref{vz3}), and $E(\sum\limits_{i,j\in[N]}|f(i,j)|^2)=\frac{N^2}{3}$, the lower bound is achieved as follows:
\begin{align}
P_{\text{success}}&=\frac{E(2|f|^2)-(\frac{N^2}{24}+\frac{N}{4})+N^2\frac{\epsilon^2\sigma^2}{2}}{E(2|f|^2)+N^2\epsilon^2\sigma^2}\nonumber\\
&=1-\frac{\frac{1}{24}+\frac{1}{4N}+\frac{\epsilon^2\sigma^2}{2}}{\frac{2}{3}+\epsilon^2\sigma^2}>1/2.
\end{align}
%Let $\delta=\displaystyle\min_{l,k}|-C_2(l)+C_1(l+1)|^2$,
\end{IEEEproof}
\emph{Remark:} For a real-world image, due to the smoothness of images, the ratio of wavelet coefficients to filter coefficients is likely to be very small, so that the probability of success of quantum denoising method is likely to be high, such as 95$\%$ shown in Figure \ref{s3}.

By treating the probability of success as a constant, we conclude that the QRT-based quantum image denoising method has a time complexity O$(\log^3 n)$, which is exponentially faster than the classical PDRT denoising method.

Also, there are more quantum techniques that allow to simulate denoising using other Daubechies wavelets \cite{fijany1998quantum} and thresholds \cite{kerenidis2017quantum2} in the quantum computation framework.

\subsection{Line detection using SIDRT}\label{va1}

\begin{figure}[!t]
\centering
\includegraphics[scale=0.45]{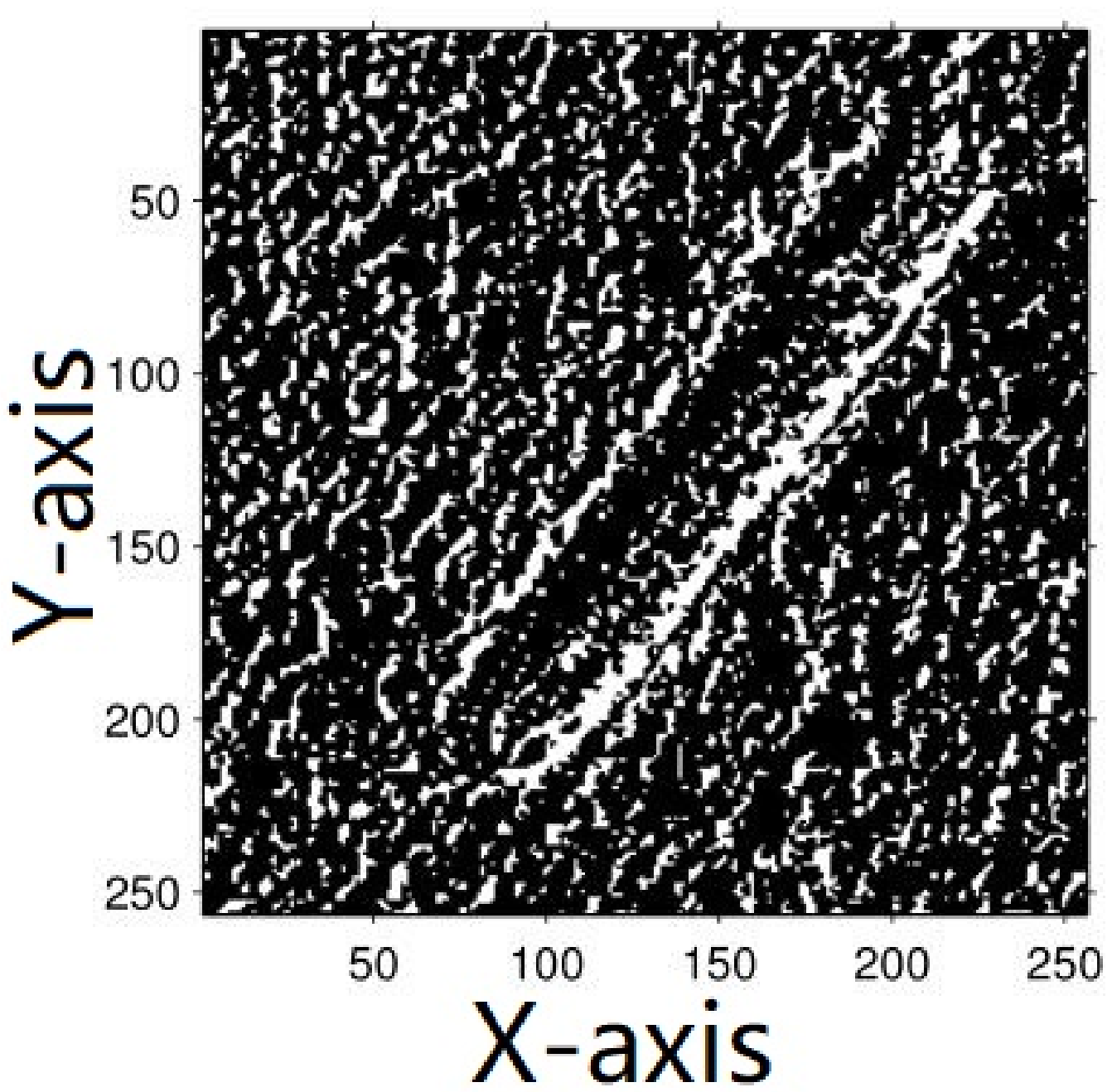}
\hfil
\includegraphics[scale=0.25]{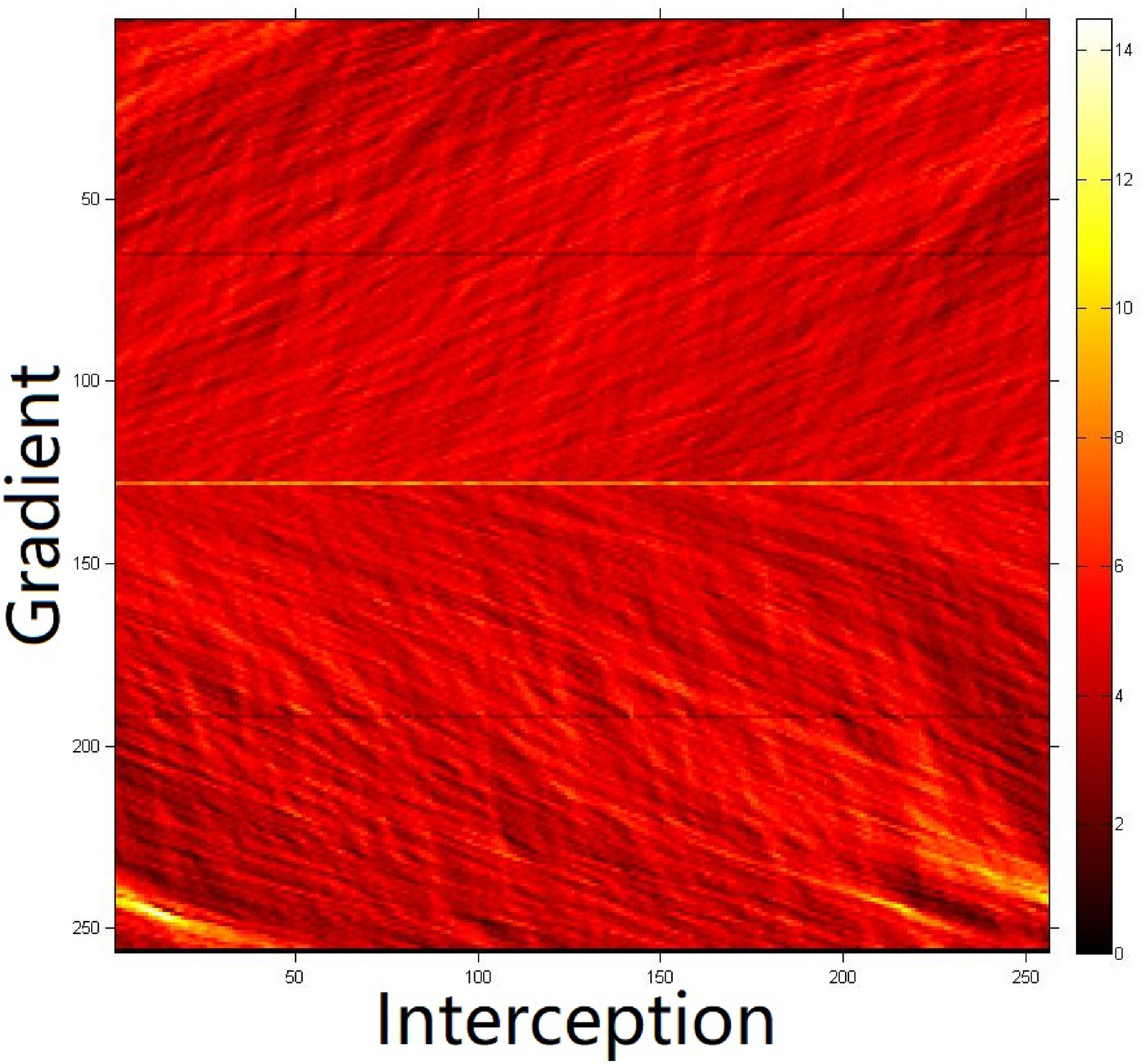}
\caption{Line detection using SIDRT. The upper is the detected image, where a distinct straight line connecting (228,53) and (97,217) is of the gradient approximately -1.251 and $y$-interception $(\frac{53}{1.251}+228 \mod  256\approx )\ 14.366$. The bottom shows the SIDRT of the upper image. The SIDRT reaches maximum at point (13,246), which implies there may be a straight line of the $y$-interception $13$ and gradient $\frac{1}{\tan (\frac{256-246}{256}-\frac{1}{4})\pi}\approx-1.284$ in the upper image. This detection result is in good agreement with the observation.}
\label{xx1}
\end{figure}

The SIDRT enables one to detect possible line in images: when the SIDRT $P_k(l)$ of image $f$ reaches maximum for some pair $l$ and $k$, there is likely to be a straight line with gradient $k$ and interception $l$ in the image.
For example, suppose that $f$ depicts a line composed of points with grayscale $1$, while the grayscales of all other points in the picture are $0$. Then the SIDRT reaches maximum at the corresponding line, with
the coordinates of the maximum in the Radon domain giving the slope and interception of the line respectively. Fig. \ref{xx1} shows a powerful line detection capability possessed by SIDRT.

%Or simply suppose that $f$ depicts a line composed of points with grayscale $1$, while the grayscales of all other points in the picture are $0$. Then the SIDRT $P_k(l)$ reaches maximum at the corresponding line, with the coordinates of the maximum in the Radon domain giving the slope and interception of the line respectively.

In classical case, performing SIDRT has a running time $\Omega(N^3)$, and finding the maximum of SIDRT has a time complexity O$(N^2)$. So, the total running time of classical SIDRT-based line detection algorithm is $\Omega(N^3)$; in contrast, there is a quantum algorithm that can run in $O(N^{1.5})$ time to execute line detections, by the following proposition:

\begin{prop}
Let $f$ be an $N\times N$ quantum image that can be prepared in time O$(T_{in})$ by the unitary $U:|0\rangle\rightarrow|\vec{f}\rangle$. There exist a quantum algorithm that can detect line in $f$ by using SIDRT in time $O(\frac{N^{1.5}T_{in}}{\epsilon}\emph{polylog N})$, where $\epsilon$ is the desired precision. This algorithm outputs a pair $(\theta_0,l_0)$ such that $\tilde{P}_{k_{\theta_0}}(l_0)\geq (1-\frac{2}{\sqrt{3}}\epsilon)\displaystyle\max_{\theta,l} P_{k_\theta}(l)$.
\end{prop}

\begin{IEEEproof}
Applying Theorem $\ref{323}$ to the input state $\sum_{\theta,l\in[N]} \frac{1}{N} |\theta \rangle |l\rangle |0\rangle$ with setting precision $\frac{\epsilon}{N^{0.5}}$, one can prepare the following state in time $O(\frac{N^{0.5}T_{in}\text{polylog N}}{\epsilon})$:
\begin{eqnarray}
\sum_{\theta,l\in[N]} \frac{1}{N} |\theta \rangle |l\rangle |\tilde{P}_{k_\theta}(l)\rangle,
\end{eqnarray}
where $|\tilde{P}_{k_\theta}(l)- P_{k_\theta}(l)|\leq \frac{\epsilon}{N^{0.5}}$. Then by the quantum algorithm for finding the maximum among $N$ terms with query complexity O$(\sqrt{N})$ \cite{ahuja1999quantum}, one can find $(l_0, \theta_0)$ such that $\tilde{P}_{k_{\theta_0}}(l_0)=\max\limits_{\theta,l \in[N]} \tilde{P}_{k_\theta}(l)$ in time O$(\frac{N^{1.5}T_{in}\text{polylog N}}{\epsilon})$.

By Proposition \ref{qwd}, in the average case, it holds that $\displaystyle\max_{\theta,l}P_{k_\theta}(l)\geq \frac{\sqrt{3}}{2N^{0.5}}$. Now that
\vspace{-0.3cm}
\begin{eqnarray}
|\tilde{P}_{k_\theta}(l)- P_{k_\theta}(l)|\leq \frac{\epsilon}{N^{0.5}}\leq\frac{2\epsilon\max P_{k_\theta}(l)}{\sqrt{3}},
\end{eqnarray}
it holds that
\begin{eqnarray}
\tilde{P}_{k_{\theta_0}}(l_0)=\max\tilde{P}_{k_\theta}(l) \geq  (1-\frac{2\epsilon}{\sqrt{3}})\max P_{k_\theta}(l).
\end{eqnarray}
\end{IEEEproof}

%As discussed previously, quantum IDRT can be achieved more quickly with the threshold being set to a higher level, because of a higher tolerable error for the higher threshold. However, if the level is set too high, that quantum thresholding method would success with a little probability. Thus, the setting of the threshold would be very delicate.
\section{Conclusion}

This paper presents a novel discrete Radon transform for efficient quantum implementation. By theoretical analysis and numerical experiments, it is shown that our new proposed quantum Radon transform has similar functionality to the classical PDRT, and is more suitable for quantum implementation than PDRT. In addition to its quantum advantage, the QRT has a classical application value that it can be used as a `reversible PDRT'.

%Our work provides the theoretical basis for using quantum computer to quickly solve DRTs-related image problems, such as denoising, line detection, etc. Also, it makes the first step for further extending DRTs-based image processing techniques, such as image encryption and watermarking, to the quantum computation framework.
Also, a polynomially fast quantum implementation of another interpolation-based kind of DRT is given. There are two problems that deserve further investigation: 1. is it possible to provide exponential speedup for performing interpolated-based DRTs or line detection? 2. Considering that the computational basis state of the maximum amplitude is most likely to be observed by quantum measurement, can this property be used to design a better line detection algorithm?

\section*{Acknowledgment}

This work is supported by Chinese Postdoctoral Science Foundation Grant No. 2020M680716, China National Key Research and Development Projects 2020YFA0712300, 2018YFA0704705, and National Natural Science Foundation of China (Grant Nos. 11471040 and 11761131002).

\ifCLASSOPTIONcaptionsoff
  \newpage
\fi



\bibliographystyle{IEEEtran}
\bibliography{IEEEabrv,apssamp}



%\vfill


\end{document}